\documentclass[aps,prx,twocolumn,floatfix,10pt,amssymb,amsfont,amsmath,superscriptaddress,float,nofootinbib,longbibliography]{revtex4-2}

\usepackage{preamb}
\pgfdeclarelayer{back}
\pgfdeclarelayer{front}
\pgfsetlayers{back,main,front}

\tikzset{->-/.style={decoration={
			markings,
			mark=at position #1 with {\arrow{Triangle[width=3.5pt, length=3.5pt]}}},postaction={decorate}}}

\tikzset{
    every picture/.append style={
        line join=round,
        line width = 0.7pt,
        line cap=round
    }
}

\tikzset{mask point/.style={ transform shape, sloped, 
minimum width=20pt, minimum height=4pt, inner sep=0pt, ultra thin, font=\tiny}}

\tikzset{
    clip even odd rule/.code={\pgfseteorule},
    invclip/.style={clip,insert path=[clip even odd rule]{
    [reset cm](-\maxdimen,-\maxdimen)rectangle(\maxdimen,\maxdimen)
    }}}

\tikzstyle{dot}=[dash pattern= on 0.6pt off 2.1pt]
\tikzstyle{obj}=[line width = 0.7pt, line join=round, line cap=round]
\tikzstyle{speObj}=[line width = 0.7pt, line join=round, line cap=round, color=BrickRed]
\tikzstyle{mod}=[color=BlueViolet,line width = 0.7pt, line join=round, line cap=round]
\tikzstyle{vac}=[draw=none]
\tikzstyle{dotmod}=[dash pattern= on 0.6pt off 2.1pt,color=BlueViolet,line width = 0.7pt, line join=round, line cap=round]
\tikzstyle{mor}=[draw=none, fill=gray!22]

\newcommand{\middleCoo}[6]{
    \path (#1) --coordinate[pos=0.5](#3) (#2);
    \path (#1) --coordinate[pos={0.5-#6}](#4) (#2);
    \path (#1) --coordinate[pos={0.5+#6}](#5) (#2);
}

\newcommand{\barycentre}[8]{
    \middleCoo{#1}{#2}{A}{}{}{#8};
    \middleCoo{#2}{#3}{B}{}{}{#8};
    \middleCoo{#1}{#3}{C}{}{}{#8};
    \path[name path=PA#3] (A) -- (#3);
    \path[name path=PB#1] (B) -- (#1);
    \path[name path=PC#2] (C) -- (#2);
    \path [name intersections={of=PA#3 and PB#1,by={#4}}];
    \path (#1) --coordinate[pos=0.9](#5) (#4);
    \path (#2) --coordinate[pos=0.9](#6) (#4);
    \path (#3) --coordinate[pos=0.9](#7) (#4);
}

\newcommand{\monoidalAssociator}[1]{
	\begin{tikzpicture}[scale=0.6,baseline={([yshift=-1ex]current bounding box.center)}]
	    \def\l{1.5};
        \ifthenelse{#1 = 1}{
            \draw[obj] (0,0) -- (0,1);
            \draw[obj] (0,1) -- (-\l,3);
            \draw[obj] (0,1) -- (\l,3);
            \draw[obj] (-\l/2,2) -- (0,3);
            \node[mor, circle, inner sep=3.3pt] at (-\l/2,2) {};
            \node at (-\l/2,2) {${\sss i}$};
            \node[mor, circle, inner sep=3.3pt] at (0,1) {};
            \node at (0,1) {${\sss k}$};
            \node at (-0.65,1.3) {${\sss Y_5}$};
        }{}
        \ifthenelse{#1 = 2}{
            \draw[obj] (0,0) -- (0,1);
            \draw[obj] (0,1) -- (-\l,3);
            \draw[obj] (0,1) -- (\l,3);
            \draw[obj] (\l/2,2) -- (0,3);
            \node[mor, circle, inner sep=3.3pt] at (0,1) {};
            \node at (0,1) {${\sss j}$};
            \node[mor, circle, inner sep=3.3pt] at (\l/2,2) {};
            \node at (\l/2,2) {${\sss l}$};
            \node at (0.7,1.3) {${\sss Y_6}$};
        }{}
        \node[above] at (-\l,3) {${\sss Y_1}$};
        \node[above] at (0,3) {${\sss Y_2}$};
        \node[above] at (\l,3) {${\sss Y_3}$};
        \node[below] at (0,0) {${\sss Y_4}$};
	\end{tikzpicture}
}

\newcommand{\PEPST}[5]{
    \def\sty{#1}
    \def\morI{#2}
    \def\morII{#3}
    \def\morIII{#4}
    \def\morIV{#5}
    \PEPSTcont
}

\newcommand{\PEPSTcont}[7]{
    \begin{tikzpicture}[scale=1.15,baseline={([yshift=-.5ex]current bounding box.center)}]
        \def\r{2pt};
        \def\l{2.8};
        \def\sp{0.05};
        
        \ifthenelse{#7 = 1 \OR #7 = 3}{\def\sgn{1}}{}
        \ifthenelse{#7 = 2}{\def\sgn{-1}}{}
        
        \coordinate (a1) at (0,0);
        \coordinate (a2) at (\l,0);
        \coordinate (a3) at (\l/2,{\sgn*sqrt(3)*\l/2});
        
        \clip (\l/7,-\l/9) rectangle (\l-\l/7,\l/2+.1);
        
        \middleCoo{a1}{a2}{m12}{m12-1}{m12-2}{\sp};
        \middleCoo{a2}{a3}{m23}{m23-2}{m23-3}{\sp};
        \middleCoo{a1}{a3}{m13}{m13-1}{m13-3}{\sp};
        \barycentre{a1}{a2}{a3}{b123}{b123-1}{b123-2}{b123-3}{\sp};
        
        \ifthenelse{#7 = 1}{
            \draw[obj,->-=0.68] (m12) --node[pos=0.28,fill=white,rectangle,inner sep=1pt] {${\sss #6}$} (b123);
            \draw[obj,->-=0.89] (b123) --node[pos=0.47,sloped,fill=white,rectangle,inner sep=0pt] {${\sss #4}$} node[pos=1.17, sloped]{${\sss \morI}$} (m13);
            \draw[obj,->-=0.89] (b123) --node[pos=0.47,sloped,fill=white,rectangle,inner sep=0pt] {${\sss #5}$} node[pos=1.17, sloped]{${\sss \morII}$} (m23);
            \node[below] at (m12) {${\sss \morIII}$}; 
        }{}
        \ifthenelse{#7 = 2}{
            \draw[obj,->-=0.7] (b123) --node[pos=0.38,fill=white,rectangle,inner sep=1pt] {${\sss #6}$} (m12);
            \draw[obj,->-=0.77] (m13) --node[pos=0.44,sloped,fill=white,rectangle,inner sep=0pt] {${\sss #4}$} (b123);
            \draw[obj,->-=0.77] (m23) --node[pos=0.44,sloped,fill=white,rectangle,inner sep=0pt] {${\sss #5}$} (b123);
        }{}
        \node[mor, regular polygon, regular polygon sides=3, inner sep=3.8pt, rounded corners=1pt] at (b123) {};
        
        \draw[rounded corners=\r, \sty] 
        (m12-1) -- (b123-1) -- (m13-1)
        (m12-2) -- (b123-2) -- (m23-2)
        (m23-3) -- (b123-3) -- (m13-3);

        \node[mor, opacity=.6, regular polygon, regular polygon sides=3, inner sep=3.8pt, rounded corners=1pt] at (b123) {};
        
        \node[yshift=-\sgn*5pt, xshift=-8pt] at (b123-1) {${\sss #1}$};
        \node[yshift=-\sgn*5pt, xshift=8pt] at (b123-2) {${\sss #2}$};
        \node[yshift=\sgn*9pt] at (b123-3) {${\sss #3}$}; 
        \node[] at (b123) {${\sss \morIV}$};
    \end{tikzpicture}
}

\newcommand{\Fmove}[9]{
    \begin{tikzpicture}[scale=1.15,baseline={([yshift=-.5ex]current bounding box.center)}]
        \def\r{2pt};
        \def\l{2.8};
        \def\sp{0.05};
        \def\w{0.23*\l};

        \coordinate (a1) at (0,0);
        \coordinate (a2) at (\l,0);
        \coordinate (a3) at (\l/2,{sqrt(3)*\l/2});
        \ifthenelse{#1=1}{
            \clip (-.09*\l,-0.08) rectangle (.78*\l,.78*\l);
            \coordinate (b1) at ($(-\l/2,{sqrt(3)*\l/2})+({\w*cos(30)},{-\w*sin(30)})$);
            \coordinate (b2) at ($(\l/2,{sqrt(3)*\l/2})+({\w*cos(30)},{-\w*sin(30)})$);
            \coordinate (b3) at ({\w*cos(30)},{-\w*sin(30)});
        }{
            \clip (\l/4-0.08,-0.08) rectangle (1.08*\l,.78*\l);
            \coordinate (b2) at ($(\l/2,{sqrt(3)*\l/2})+({-\w*cos(30)},{-\w*sin(30)})$);
            \coordinate (b1) at ($(\l+\l/2,{sqrt(3)*\l/2})+({-\w*cos(30)},{-\w*sin(30)})$);
            \coordinate (b3) at ({\l-\w*cos(30)},{-\w*sin(30)});
        }
        
        \middleCoo{a1}{a2}{m12}{m12-1}{m12-2}{\sp};
        \middleCoo{a2}{a3}{m23}{m23-2}{m23-3}{\sp};
        \middleCoo{a1}{a3}{m13}{m13-1}{m13-3}{\sp};
        \barycentre{a1}{a2}{a3}{b123}{b123-1}{b123-2}{b123-3}{\sp};
        \middleCoo{b1}{b2}{n12}{n12-1}{n12-2}{\sp};
        \middleCoo{b2}{b3}{n23}{n23-2}{n23-3}{\sp};
        \middleCoo{b1}{b3}{n13}{n13-1}{n13-3}{\sp};
        \barycentre{b1}{b2}{b3}{c123}{c123-1}{c123-2}{c123-3}{\sp};
        
        \draw[obj,->-=0.68] (m12) --node[pos=0.28,fill=white,rectangle,inner sep=1pt] {${\sss #5}$} (b123);
        \draw[obj,->-=0.9] (b123) --node[pos=0.48,sloped,fill=white,rectangle,inner sep=0pt] {${\sss #6}$} (m13);
        \draw[obj,->-=0.89] (b123) --node[pos=0.47,sloped,fill=white,rectangle,inner sep=0pt] {${\sss #4}$} (m23);
        \draw[obj,->-=0.89] (c123) --node[pos=0.47,fill=white,rectangle,inner sep=1pt] {${\sss #3}$} (n12);
        \draw[obj,->-=0.89] (c123) --node[pos=0.47,sloped,fill=white,rectangle,inner sep=0pt] {${\sss #2}$} (n13);
      
        \node[mor, regular polygon, regular polygon sides=3, inner sep=3.8pt, rounded corners=1pt] at (b123) {};
        \node at (b123) {${\sss #8}$};
        \node[mor, regular polygon, regular polygon sides=3, inner sep=3.8pt, rounded corners=1pt, rotate=180] at (c123) {};
        \node at (c123) {${\sss #7}$};

        \ifthenelse{#1=1}{
            \draw[#9, rounded corners=\r] (m12-1) -- (b123-1) -- (c123-3) -- (n13-3);
            \draw[#9, rounded corners=\r] (m12-2) -- (b123-2) -- (m23-2);
            \draw[#9, rounded corners=\r] (m23-3) -- (b123-3) -- (c123-2) -- (n12-2);
            \draw[#9, rounded corners=\r] (n12-1) -- (c123-1) -- (n13-1);
        }{
            \draw[#9, rounded corners=\r] (m12-2) -- (b123-2) -- (c123-3) -- (n13-3);
            \draw[#9, rounded corners=\r] (m12-1) -- (b123-1) -- (m13-1);
            \draw[#9, rounded corners=\r] (m13-3) -- (b123-3) -- (c123-2) -- (n12-2);
            \draw[#9, rounded corners=\r] (n12-1) -- (c123-1) -- (n13-1);
        }

        \node[mor, opacity=.6, regular polygon, regular polygon sides=3, inner sep=3.8pt, rounded corners=1pt] at (b123) {};
        \node at (b123) {${\sss #8}$};
        \node[mor, opacity=.6, regular polygon, regular polygon sides=3, inner sep=3.8pt, rounded corners=1pt, rotate=180] at (c123) {};
        \node at (c123) {${\sss #7}$};
    \end{tikzpicture}
}

\newcommand{\hamOp}[8]{
    \begin{tikzpicture}[scale=1.15,baseline={([yshift=-.5ex]current bounding box.center)}]
        \def\ll{2.8};
        \def\l{.808};
        \def\h{1.3};
        \def\sp{0.05};
        \def\r{2pt};

        \draw[obj,->-=0.75] (0,\l) --node[pos=0.4,fill=white,rectangle,inner sep=1pt] {${\sss #5}$} (\h*\l,\l);
        \draw[obj,->-=0.68] (0,0) --node[pos=0.28,fill=white,rectangle,inner sep=1pt] {${\sss #3}$} (0,\l);
        \draw[obj,->-=0.89] (0,\l) --node[pos=0.49,fill=white,rectangle,inner sep=1pt] {${\sss #1}$} (0,2*\l);
        \node[mor, regular polygon, regular polygon sides=3, inner sep=3.9pt, rounded corners=1pt, rotate=-30] at (.04,\l) {};
        \draw[obj,->-=0.68] (\h*\l,0) --node[pos=0.28,fill=white,rectangle,inner sep=1pt] {${\sss #4}$} (\h*\l,\l);
        \draw[obj,->-=0.89] (\h*\l,\l) --node[pos=0.49,fill=white,rectangle,inner sep=1pt] {${\sss #2}$} (\h*\l,2*\l);
        \node[mor, regular polygon, regular polygon sides=3, inner sep=3.9pt, rounded corners=1pt, rotate=30] at (\h*\l-.04,\l) {};

        \ifthenelse{#8=2}{
            \draw[mod] (-\sp*\ll,0) -- (-\sp*\ll,2*\l);
            \draw[mod, rounded corners=\r] (\sp*\ll,0) -- (\sp*\ll,\l-\sp*\ll) -- (\h*\l-\sp*\ll,\l-\sp*\ll) -- (\h*\l-\sp*\ll,0);
            \draw[mod] (\h*\l+\sp*\ll,0) -- (\h*\l+\sp*\ll,2*\l);
            \draw[mod, rounded corners=\r] (\sp*\ll,2*\l) -- (\sp*\ll,\l+\sp*\ll) -- (\h*\l-\sp*\ll,\l+\sp*\ll) -- (\h*\l-\sp*\ll,2*\l);
        }{}
        \node[mor, opacity=.6, regular polygon, regular polygon sides=3, inner sep=3.9pt, rounded corners=1pt, rotate=-30] at (.04,\l) {};
        \node[mor, opacity=.6, regular polygon, regular polygon sides=3, inner sep=3.9pt, rounded corners=1pt, rotate=30] at (\h*\l-.04,\l) {};
        \node at (0,\l) {${\sss #6}$};
        \node at (\h*\l,\l) {${\sss #7}$};
    \end{tikzpicture}
}

\newcommand{\MPO}[0]{
    \begin{tikzpicture}[scale=1.15,baseline={([yshift=-.5ex]current bounding box.center)}]
        \def\ll{2.8};
        \def\l{.808};
        \def\h{1.3};
        \def\sp{0.05};
        \def\r{2pt};    

        \draw[obj,->-=0.92] (0,0) --node[pos=0.53,fill=white,rectangle,inner sep=1pt] {${\sss \ub 3}$} (0,\l);
        \draw[obj,->-=0.66] (0,-\l) --node[pos=0.26,fill=white,rectangle,inner sep=1pt] {${\sss \ub 3}$} (0,0);
        \node[mor, rectangle, inner sep=7.4pt, rounded corners=1pt] at (0,0) {};        
        \draw[mod, rounded corners=\r] (-\sp*\ll,-\l) -- (-\sp*\ll,-\sp*\ll) -- (-\l,-\sp*\ll);
        \draw[mod, rounded corners=\r] (\sp*\ll,-\l) -- (\sp*\ll,-\sp*\ll) -- (\l,-\sp*\ll);

        \node[mor, rectangle, opacity=.6,  inner sep=7.4pt, rounded corners=1pt] at (0,0) {}; 
        \node[above] at (0,\l) {${\sss v}$};
        \node[left, xshift=3pt] at (-\l,0) {${\sss (V_1,v_1)}$};
        \node[right, xshift=-3pt] at (\l,0) {${\sss (V_2,v_2)}$};
        \node[below] at (0,-\l) {${\sss i}$};
    \end{tikzpicture}
}

\newcommand{\symTN}[1]{
    \begin{tikzpicture}[scale=1.15,baseline={([yshift=-.5ex]current bounding box.center)}]
        \def\ll{2.8};
        \def\l{.808};
        \def\h{1.3};
        \def\sp{0.05};
        \def\r{2pt};    

        \ifthenelse{#1=1}{
            \draw[obj,->-=0.92] (0,0) --node[pos=0.53,fill=white,rectangle,inner sep=1pt] {${\sss \ub 3}$} (0,\l);
            \draw[obj,->-=.7] (\l,0) --node[pos=0.28,fill=white,rectangle,inner sep=.5pt](){${\sss V_2}$} (0,0);
            \draw[obj,->-=.95] (0,0) --node[pos=0.53,fill=white,rectangle,inner sep=.5pt](){${\sss V_1}$} (-\l,0);
    
            \node[mor, regular polygon, regular polygon sides=3, inner sep=3.8pt, rounded corners=1pt, rotate=60] at (0,0) {};
            \node at (0,0) {${\sss i}$};
            \node[right] at (\l,0) {${\sss v_2}$};
            \node[left] at (-\l,0) {${\sss v_1}$};
            \node[above] at (0,\l) {${\sss v}$};
        }{}

        \ifthenelse{#1=0}{
            \draw[obj,->-=0.92] (0,0) --node[pos=0.53,fill=white,rectangle,inner sep=1pt] {${\sss \ub 3}$} (0,\l);
            \draw[obj] (\l,0) -- (-\l,0);
            \node[mor, rectangle, inner sep=7.4pt, rounded corners=1pt] at (0,0) {};
            \node[above] at (0,\l) {${\sss v}$};
            \node[left] at (-\l,0) {${\sss u_1}$};
            \node[right] at (\l,0) {${\sss u_2}$};
        }{}
        \ifthenelse{#1=2}{
            \draw[obj] (\l,0) -- (-\l,0);
            \node[mor, rectangle, inner sep=7.4pt, rounded corners=1pt] at (0,0) {};
            \node[] at (0,0) {${\sss i}$};
            \node[left] at (-\l,0) {${\sss (V_1,d_1)}$};
            \node[right] at (\l,0) {${\sss (V_2,d_2)}$};
        }{}
        
    \end{tikzpicture}
}

\newcommand{\MPSTen}[1]{
    \begin{tikzpicture}[scale=1.15,baseline={([yshift=-.5ex]current bounding box.center)}]
        \def\ll{2.8};
        \def\l{.808};
        \def\h{1.3};
        \def\sp{0.05};
        \def\r{2pt};    

        \draw[obj,->-=0.92] (0,0) --node[pos=0.53,fill=white,rectangle,inner sep=1pt] {${\sss \ub 3}$} (0,\l);
        \draw[obj] (\l,0) -- (-\l,0);
        \node[mor, rectangle, inner sep=7.4pt, rounded corners=1pt] at (0,0) {};
        \draw[mod, rounded corners=\r] (-\l,\sp*\ll) -- (-\sp*\ll,\sp*\ll) -- (-\sp*\ll,\l); 
        \draw[mod, rounded corners=\r] (\l,\sp*\ll) -- (\sp*\ll,\sp*\ll) -- (\sp*\ll,\l); 
        \node[mor, rectangle, opacity=.6, inner sep=7.4pt, rounded corners=1pt] at (0,0) {};
        \node[] at (0,0) {${\sss A}$};
        
        \ifthenelse{#1=1}{
            \node[left] at (-\l,0) {${\sss (V_1,d_1)}$};
            \node[right] at (\l,0) {${\sss (V_2,d_2)}$};
            \node[above] at (0,\l) {${\sss (V_1 \ub 3 V_2, i)}$};
        }{}
        
    \end{tikzpicture}
}

\newcommand{\MPSDoubleTen}[1]{
    \begin{tikzpicture}[scale=1.15,baseline={([yshift=-.5ex]current bounding box.center)}]
        \def\ll{2.8};
        \def\l{.808};
        \def\h{1.3};
        \def\sp{0.05};
        \def\r{2pt};    

        \draw[obj,->-=0.92] (0,0) --node[pos=0.53,fill=white,rectangle,inner sep=1pt] {${\sss \ub 3}$} (0,\l);
        \draw[obj,->-=0.92] (\h*\l,0) --node[pos=0.53,fill=white,rectangle,inner sep=1pt] {${\sss \ub 3}$} (\h*\l,\l);
        \draw[obj] (-\l,0) -- (\h*\l+\l,0);    
        \node[left] at (-\l,0) {${\sss (V_1,d_1)}$};
        \node[right] at (\h*\l+\l,0) {${\sss (V_3,d_3)}$};
        \node[above, xshift=-3pt] at (0,\l) {${\sss (V_1 \ub 3 V_2, i_1)}$};
        \node[above, xshift=3pt] at (\h*\l,\l) {${\sss (V_2 \ub 3 V_3, i_2)}$};

        \ifthenelse{#1=1}{
            \node[mor, rectangle, inner sep=7.4pt, rounded corners=1pt, minimum width=49pt] at (.5*\h*\l,0) {};
            \draw[mod, rounded corners=\r] (-\l,\sp*\ll) -- (-\sp*\ll,\sp*\ll) -- (-\sp*\ll,\l); 
            \draw[mod, rounded corners=\r] (\sp*\ll,\l) -- (\sp*\ll,\sp*\ll) -- (\h*\l-\sp*\ll,\sp*\ll) -- (\h*\l-\sp*\ll,\l);
            \draw[mod, rounded corners=\r] (\h*\l+\sp*\ll,\l) -- (\h*\l+\sp*\ll,\sp*\ll) -- (\h*\l+\l,\sp*\ll);
            \node[mor, opacity=.6, rectangle, inner sep=7.4pt, rounded corners=1pt, minimum width=49pt] at (.5*\h*\l,0) {};
            \node[] at (.5*\h*\l,0) {${\sss B}$};
        }{}
        \ifthenelse{#1=2}{
            \node[mor, rectangle, inner sep=7.4pt, rounded corners=1pt] at (0,0) {};
            \node[mor, rectangle, inner sep=7.4pt, rounded corners=1pt] at (\h*\l,0) {};
            \draw[mod, rounded corners=\r] (-\l,\sp*\ll) -- (-\sp*\ll,\sp*\ll) -- (-\sp*\ll,\l); 
            \draw[mod, rounded corners=\r] (\sp*\ll,\l) -- (\sp*\ll,\sp*\ll) -- (\h*\l-\sp*\ll,\sp*\ll) -- (\h*\l-\sp*\ll,\l);
            \draw[mod, rounded corners=\r] (\h*\l+\sp*\ll,\l) -- (\h*\l+\sp*\ll,\sp*\ll) -- (\h*\l+\l,\sp*\ll);
            \node[mor, opacity=.6, rectangle, inner sep=7.4pt, rounded corners=1pt] at (0,0) {};
            \node[mor, opacity=.6, rectangle, inner sep=7.4pt, rounded corners=1pt] at (\h*\l,0) {};
            \node[] at (0,0) {${\sss A_\msf i}$};
            \node[] at (\h*\l,0) {${\sss A_{\msf i+1}}$};
        }{}
    
    \end{tikzpicture}
}

\newcommand{\dualMPS}[0]{
    \begin{tikzpicture}[scale=1.15,baseline={([yshift=-.5ex]current bounding box.center)}]
        \def\ll{2.8};
        \def\l{.808};
        \def\h{1.3};
        \def\sp{0.05};
        \def\r{2pt};

        \draw[obj] (-.5*\h*\l,0) -- (2.5*\h*\l,0);
        \draw[obj, line width=.9pt, dash pattern= on 0pt off 3pt] (-.59*\h*\l,0) -- (-.77*\h*\l,0);
        \draw[obj, line width=.9pt, dash pattern= on 0pt off 3pt] (2.59*\h*\l,0) -- (2.77*\h*\l,0);
        
        \draw[mod, line width=.9pt, dash pattern= on 0pt off 3pt] (-.59*\h*\l,\sp*\ll) -- (-.77*\h*\l,\sp*\ll);
        \draw[mod, line width=.9pt, dash pattern= on 0pt off 3pt] (-.59*\h*\l,\h*\l-\sp*\ll) -- (-.77*\h*\l,\h*\l-\sp*\ll);
        \draw[mod, line width=.9pt, dash pattern= on 0pt off 3pt] (2.59*\h*\l,\sp*\ll) -- (2.77*\h*\l,\sp*\ll);
        \draw[mod, line width=.9pt, dash pattern= on 0pt off 3pt] (2.59*\h*\l,\h*\l-\sp*\ll) -- (2.77*\h*\l,\h*\l-\sp*\ll);
        
        \foreach \x in {0,1,2}{
            \draw[obj,->-=0.73] (\x*\h*\l,0) --node[pos=0.41,fill=white,rectangle,inner sep=1pt] {${\sss \ub 3}$} (\x*\h*\l,\h*\l);
            \draw[obj,->-=0.92] (\x*\h*\l,\h*\l) --node[pos=0.53,fill=white,rectangle,inner sep=1pt] {${\sss \ub 3}$} (\x*\h*\l,\h*\l+\l);
            \node[mor, rectangle, inner sep=7.4pt, rounded corners=1pt] at (\x*\h*\l,0) {};
            \node[mor, rectangle, inner sep=7.4pt, rounded corners=1pt] at (\x*\h*\l,\h*\l) {};
            \draw[mod, rounded corners=\r] (\x*\h*\l-\h/2*\l,\h*\l-\sp*\ll) -- (\x*\h*\l-\sp*\ll,\h*\l-\sp*\ll) -- (\x*\h*\l-\sp*\ll, \sp*\ll) -- (\x*\h*\l-\h/2*\l,\sp*\ll);
            \draw[mod, rounded corners=\r] (\x*\h*\l+\h/2*\l,\h*\l-\sp*\ll) -- (\x*\h*\l+\sp*\ll,\h*\l-\sp*\ll) -- (\x*\h*\l+\sp*\ll, \sp*\ll) -- (\x*\h*\l+\h/2*\l,\sp*\ll);
            \node[mor, opacity=.6, rectangle, inner sep=7.4pt, rounded corners=1pt] at (\x*\h*\l,0) {};
            \node[mor, opacity=.6, rectangle, inner sep=7.4pt, rounded corners=1pt] at (\x*\h*\l,\h*\l) {};
        }

        \node at (0,0) {${\sss A_\msf i-1}$};
        \node at (\h*\l,0) {${\sss A_{\msf i}}$};
        \node at (2*\h*\l,0) {${\sss A_{\msf i+1}}$};
    \end{tikzpicture}
}

\newcommand{\symMPO}[0]{
    \begin{tikzpicture}[scale=1.15,baseline={([yshift=-0.5ex]current bounding box.center)}]
        \def\ll{2.8};
        \def\l{.808};
        \def\h{1.3};
        \def\sp{0.05};
        \def\r{2pt};
        
        \draw[speObj, line width=.9pt, dash pattern= on 0pt off 3pt] (-.59*\h*\l,\h*\l) -- (-.77*\h*\l,\h*\l);
        \draw[mod, line width=.9pt, dash pattern= on 0pt off 3pt] (-.59*\h*\l,\h*\l-\sp*\ll) -- (-.77*\h*\l,\h*\l-\sp*\ll);
        \draw[mod, line width=.9pt, dash pattern= on 0pt off 3pt] (-.59*\h*\l,\h*\l+\sp*\ll) -- (-.77*\h*\l,\h*\l+\sp*\ll);

        \draw[speObj, line width=.9pt, dash pattern= on 0pt off 3pt] (2.59*\h*\l,\h*\l) -- (2.77*\h*\l,\h*\l);
        \draw[mod, line width=.9pt, dash pattern= on 0pt off 3pt] (2.59*\h*\l,\h*\l-\sp*\ll) -- (2.77*\h*\l,\h*\l-\sp*\ll);
        \draw[mod, line width=.9pt, dash pattern= on 0pt off 3pt] (2.59*\h*\l,\h*\l+\sp*\ll) -- (2.77*\h*\l,\h*\l+\sp*\ll);
        
        \foreach \x in {0,1,2}{
            \draw[obj,->-=0.45] (\x*\h*\l,\h*\l-\l) -- (\x*\h*\l,\h*\l);
            \draw[obj,->-=0.73] (\x*\h*\l,\h*\l) -- (\x*\h*\l,\h*\l+\l);
            \node[mor, rectangle, inner sep=7.4pt, rounded corners=1pt] at (\x*\h*\l,\h*\l) {};
            \draw[speObj] (\x*\h*\l-.5*\h*\l,\h*\l) -- (\x*\h*\l+.5*\h*\l,\h*\l);

            \draw[mod, rounded corners=\r] (\x*\h*\l-\sp*\ll,\h*\l-\l) -- (\x*\h*\l-\sp*\ll,\h*\l-\sp*\ll) -- (\x*\h*\l-\h/2*\l,\h*\l-\sp*\ll);
            \draw[mod, rounded corners=\r] (\x*\h*\l+\sp*\ll,\h*\l-\l) -- (\x*\h*\l+\sp*\ll,\h*\l-\sp*\ll) -- (\x*\h*\l+\h/2*\l,\h*\l-\sp*\ll);
            
            \draw[mod, rounded corners=\r] (\x*\h*\l-\sp*\ll,\h*\l+\l) -- (\x*\h*\l-\sp*\ll,\h*\l+\sp*\ll) -- (\x*\h*\l-\h/2*\l,\h*\l+\sp*\ll);
            \draw[mod, rounded corners=\r] (\x*\h*\l+\sp*\ll,\h*\l+\l) -- (\x*\h*\l+\sp*\ll,\h*\l+\sp*\ll) -- (\x*\h*\l+\h/2*\l,\h*\l+\sp*\ll);

            \node[mor, opacity=.6, rectangle, inner sep=7.4pt, rounded corners=1pt] at (\x*\h*\l,\h*\l) {};
        }
        \foreach \x in {0,1,2,3}{
            \node[mod] at (\x*\h*\l-\h/2*\l,\h*\l/2) {${\sss \mc Q}$};
            \node[mod] at (\x*\h*\l-\h/2*\l,3*\h*\l/2) {${\sss \mc Q}$};
        }
        \node[speObj, xshift=-15pt] at (-\h*\l,\h*\l) {${\sss \mc C^*_\mc P \simeq {(\mc C^*_\mc R)}^*_\mc Q}$};
    \end{tikzpicture}
}

\newcommand{\dualMPSG}[0]{
    \begin{tikzpicture}[scale=1.15,baseline={([yshift=-0.5ex]current bounding box.center)}]
        \def\ll{2.8};
        \def\l{.808};
        \def\h{1.3};
        \def\sp{0.05};
        \def\r{2pt};

        \draw[obj] (-.5*\h*\l,0) -- (2.5*\h*\l,0);
        \draw[obj, line width=.9pt, dash pattern= on 0pt off 3pt] (-.59*\h*\l,0) -- (-.77*\h*\l,0);
        \draw[obj, line width=.9pt, dash pattern= on 0pt off 3pt] (2.59*\h*\l,0) -- (2.77*\h*\l,0);
        
        \draw[speObj, line width=.9pt, dash pattern= on 0pt off 3pt] (-.59*\h*\l,\h*\l) -- (-.77*\h*\l,\h*\l);
        \draw[mod, line width=.9pt, dash pattern= on 0pt off 3pt] (-.59*\h*\l,\sp*\ll) -- (-.77*\h*\l,\sp*\ll);
        \draw[mod, line width=.9pt, dash pattern= on 0pt off 3pt] (-.59*\h*\l,\h*\l-\sp*\ll) -- (-.77*\h*\l,\h*\l-\sp*\ll);
        \draw[mod, line width=.9pt, dash pattern= on 0pt off 3pt] (-.59*\h*\l,\h*\l+\sp*\ll) -- (-.77*\h*\l,\h*\l+\sp*\ll);

        \draw[speObj, line width=.9pt, dash pattern= on 0pt off 3pt] (2.59*\h*\l,\h*\l) -- (2.77*\h*\l,\h*\l);
        \draw[mod, line width=.9pt, dash pattern= on 0pt off 3pt] (2.59*\h*\l,\sp*\ll) -- (2.77*\h*\l,\sp*\ll);
        \draw[mod, line width=.9pt, dash pattern= on 0pt off 3pt] (2.59*\h*\l,\h*\l-\sp*\ll) -- (2.77*\h*\l,\h*\l-\sp*\ll);
        \draw[mod, line width=.9pt, dash pattern= on 0pt off 3pt] (2.59*\h*\l,\h*\l+\sp*\ll) -- (2.77*\h*\l,\h*\l+\sp*\ll);
        
        \foreach \x in {0,1,2}{
            \draw[obj,->-=0.55] (\x*\h*\l,0) -- (\x*\h*\l,\h*\l);
            \draw[obj,->-=0.73] (\x*\h*\l,\h*\l) -- (\x*\h*\l,\h*\l+\l);
            \node[mor, rectangle, inner sep=7.4pt, rounded corners=1pt] at (\x*\h*\l,0) {};
            \node[mor, rectangle, inner sep=7.4pt, rounded corners=1pt] at (\x*\h*\l,\h*\l) {};
            \draw[speObj] (\x*\h*\l-.5*\h*\l,\h*\l) -- (\x*\h*\l+.5*\h*\l,\h*\l);

            \draw[mod, rounded corners=\r] (\x*\h*\l-\h/2*\l,\h*\l-\sp*\ll) -- (\x*\h*\l-\sp*\ll,\h*\l-\sp*\ll) -- (\x*\h*\l-\sp*\ll, \sp*\ll) -- (\x*\h*\l-\h/2*\l,\sp*\ll);
            \draw[mod, rounded corners=\r] (\x*\h*\l+\h/2*\l,\h*\l-\sp*\ll) -- (\x*\h*\l+\sp*\ll,\h*\l-\sp*\ll) -- (\x*\h*\l+\sp*\ll, \sp*\ll) -- (\x*\h*\l+\h/2*\l,\sp*\ll);
            
            \draw[mod, rounded corners=\r] (\x*\h*\l-\sp*\ll,\h*\l+\l) -- (\x*\h*\l-\sp*\ll,\h*\l+\sp*\ll) -- (\x*\h*\l-\h/2*\l,\h*\l+\sp*\ll);
            \draw[mod, rounded corners=\r] (\x*\h*\l+\sp*\ll,\h*\l+\l) -- (\x*\h*\l+\sp*\ll,\h*\l+\sp*\ll) -- (\x*\h*\l+\h/2*\l,\h*\l+\sp*\ll);

            \node[mor, opacity=.6, rectangle, inner sep=7.4pt, rounded corners=1pt] at (\x*\h*\l,0) {};
            \node[mor, opacity=.6, rectangle, inner sep=7.4pt, rounded corners=1pt] at (\x*\h*\l,\h*\l) {};
        }
        \foreach \x in {0,1,2,3}{
            \node[mod] at (\x*\h*\l-\h/2*\l,\h*\l/2) {$\sss \mc Q$};
            \node[mod] at (\x*\h*\l-\h/2*\l,3*\h*\l/2) {$\sss \mc R$};
        }
        \node at (0,0) {${\sss A_{\msf i-1}}$};
        \node at (\h*\l,0) {${\sss A_\msf i}$};
        \node at (2*\h*\l,0) {${\sss A_{\msf i+1}}$};
        \node[speObj, xshift=-17pt] at (-\h*\l,\h*\l) {$\sss \Fun_{\mc C^*_\mc R}\!(\mc Q,\mc R)$};
    \end{tikzpicture}
}

\newcommand{\densityMat}[1]{
    \begin{tikzpicture}[scale=1.15,baseline={([yshift=-.5ex]current bounding box.center)}]
        \def\ll{2.8};
        \def\l{.808};
        \def\h{1.3};
        \def\r{2pt};
        \def\sp{0.05};

        \ifthenelse{#1=1}{
            \draw[obj, rounded corners=\r] (0,0) -- (.5*\h*\l,0) -- (.5*\h*\l,\h*\l) -- (0,\h*\l);
            \node[mor, rectangle, inner sep=7.4pt, rounded corners=1pt] at (.5*\h*\l,.5*\h*\l) {};
            \draw[mod, rounded corners=\r] (0,\sp*\ll) -- (.5*\h*\l-\sp*\ll,\sp*\ll) -- (.5*\h*\l-\sp*\ll,\h*\l-\sp*\ll) -- (0,\h*\l-\sp*\ll);
            \node[mor, opacity=.6, rectangle, inner sep=7.4pt, rounded corners=1pt] at (.5*\h*\l,.5*\h*\l) {};
            \node[] at (.5*\h*\l,.5*\h*\l) {${\sss \rho_\msf i}$};
        }{}
        
        \ifthenelse{#1=2}{
            \foreach \y in {0,1}{
                \ifthenelse{\y=0}{
                    \def\sgn{+1};
                }{
                    \def\sgn{-1};
                }
                \draw[mod, line width=.9pt, dash pattern= on 0pt off 3pt] (2.59*\h*\l,\y*\h*\l+\sgn*\sp*\ll) -- (2.77*\h*\l,\y*\h*\l+\sgn*\sp*\ll);
                \draw[obj] (-.5*\h*\l,\y*\h*\l) -- (2.5*\h*\l,\y*\h*\l);  
                \draw[obj, line width=.9pt, dash pattern= on 0pt off 3pt] (2.59*\h*\l,\y*\h*\l) -- (2.77*\h*\l,\y*\h*\l);
            }
            \foreach \x in {0,1,2}{
                \draw[obj,->-=0.73] (\x*\h*\l,0) --node[pos=0.41,fill=white,rectangle,inner sep=1pt] {${\sss \ub 3}$} (\x*\h*\l,\h*\l);
                \node[mor, rectangle, inner sep=7.4pt, rounded corners=1pt] at (\x*\h*\l,0) {};
                \node[mor, rectangle, inner sep=7.4pt, rounded corners=1pt] at (\x*\h*\l,\h*\l) {};
                \draw[mod, rounded corners=\r] (\x*\h*\l-.5*\h*\l,\sp*\ll) -- (\x*\h*\l-\sp*\ll,\sp*\ll) -- (\x*\h*\l-\sp*\ll,\h*\l-\sp*\ll) -- (\x*\h*\l-.5*\h*\l,\h*\l-\sp*\ll);
                \draw[mod, rounded corners=\r] (\x*\h*\l+.5*\h*\l,\sp*\ll) -- (\x*\h*\l+\sp*\ll,\sp*\ll) -- (\x*\h*\l+\sp*\ll,\h*\l-\sp*\ll) -- (\x*\h*\l+.5*\h*\l,\h*\l-\sp*\ll);
                \node[mor, opacity=.6, rectangle, inner sep=7.4pt, rounded corners=1pt] at (\x*\h*\l,0) {};
                \node[mor, opacity=.6, rectangle, inner sep=7.4pt, rounded corners=1pt] at (\x*\h*\l,\h*\l) {};
            }
            \node at (0,0) {${\sss A_{\msf i}}$};
            \node at (\h*\l,0) {${\sss A_{\msf i+1}}$};
            \node at (2*\h*\l,0) {${\sss A_{\msf i+2}}$};
            \node at (0,\h*\l) {${\sss \bar A_{\msf i}}$};
            \node at (\h*\l,\h*\l) {${\sss \bar A_{\msf i+1}}$};
            \node at (2*\h*\l,\h*\l) {${\sss \bar A_{\msf i+2}}$};
        }{}

        \ifthenelse{#1=3}{
            \draw[obj, rounded corners=\r] (.5*\h*\l,0) -- (-.5*\h*\l,0) -- (-.5*\h*\l,\h*\l) -- (.5*\h*\l,\h*\l);
            \draw[obj,->-=0.73] (0,0) --node[pos=0.41,fill=white,rectangle,inner sep=1pt] {${\sss \ub 3}$} (0,\h*\l);
            \node[mor, rectangle, inner sep=7.4pt, rounded corners=1pt] at (0,0) {};
            \node[mor, rectangle, inner sep=7.4pt, rounded corners=1pt] at (0,\h*\l) {};
            \draw[mod, rounded corners=\r] (-\sp*\ll,.5*\h*\l) -- (-\sp*\ll,\h*\l-\sp*\ll) -- (-.5*\h*\l+\sp*\ll,\h*\l-\sp*\ll) -- (-.5*\h*\l+\sp*\ll,\sp*\ll) -- (-\sp*\ll,\sp*\ll) --cycle;
            \draw[mod, rounded corners=\r] (.5*\h*\l,\sp*\ll) -- (\sp*\ll,\sp*\ll) -- (\sp*\ll,\h*\l-\sp*\ll) -- (.5*\h*\l,\h*\l-\sp*\ll);
            \node[mor, opacity=.6, rectangle, inner sep=7.4pt, rounded corners=1pt] at (0,0) {};
            \node[mor, opacity=.6, rectangle, inner sep=7.4pt, rounded corners=1pt] at (0,\h*\l) {};
            \node at (0,0) {${\sss A_\msf i}$};
            \node at (0,\h*\l) {${\sss \bar A_\msf i}$};
        }{}

        \ifthenelse{#1=4}{
            \draw[obj, rounded corners=\r] (.5*\h*\l,0) -- (-.5*\h*\l,0) -- (-.5*\h*\l,\h*\l) -- (.5*\h*\l,\h*\l);
            \draw[mod, rounded corners=\r] (.5*\h*\l,\sp*\ll) -- (-.5*\h*\l+\sp*\ll,\sp*\ll) -- (-.5*\h*\l+\sp*\ll,\h*\l-\sp*\ll) -- (.5*\h*\l,\h*\l-\sp*\ll);
        }{}
    \end{tikzpicture}
}

\newcommand{\inter}[8]{
    \begin{tikzpicture}[scale=1.15,baseline={([yshift=-.5ex]current bounding box.center)}]
        \def\ll{2.8};
        \def\l{.808};
        \def\h{1.3};
        \def\r{2pt};
        \def\sp{0.05};

        \ifthenelse{#8=1}{
            \draw[obj,->-=0.75] (0,\l) --node[pos=0.4,fill=white,rectangle,inner sep=1pt] {${\sss #5}$} (\h*\l,\l);
            \draw[obj,->-=0.68] (0,0) --node[pos=0.28,fill=white,rectangle,inner sep=1pt] {${\sss #3}$} (0,\l);
            \draw[obj,->-=0.89] (0,\l) --node[pos=0.49,fill=white,rectangle,inner sep=1pt] {${\sss #1}$} (0,2*\l);
            \draw[obj,->-=0.68] (\h*\l,0) --node[pos=0.28,fill=white,rectangle,inner sep=1pt] {${\sss #4}$} (\h*\l,\l);
            \draw[obj,->-=0.89] (\h*\l,\l) --node[pos=0.49,fill=white,rectangle,inner sep=1pt] {${\sss #2}$} (\h*\l,2*\l);
            \draw[obj,->-=.91] (\h*\l,2.3*\l) --node[pos=0.54,fill=white,rectangle,inner sep=1pt] {${\sss #1}$}  (\h*\l,2.3*\l+\l);
            \draw[obj,->-=.91] (0,2.3*\l) --node[pos=0.54,fill=white,rectangle,inner sep=1pt] {${\sss #1}$}  (0,2.3*\l+\l);
            \node[mor, rectangle, inner sep=7.4pt, rounded corners=1pt] at (0,2.3*\l) {};
            \node[mor, rectangle, inner sep=7.4pt, rounded corners=1pt] at (\h*\l,2.3*\l) {};
            \node[mor, regular polygon, regular polygon sides=3, inner sep=3.9pt, rounded corners=1pt, rotate=30] at (\h*\l-.04,\l) {};
            \node[mor, regular polygon, regular polygon sides=3, inner sep=3.9pt, rounded corners=1pt, rotate=-30] at (.04,\l) {};
            \draw[mod, rounded corners=\r] (-\sp*\ll,0) -- (-\sp*\ll,2.3*\l-\sp*\ll) -- (-.5*\h*\l,2.3*\l-\sp*\ll);
            \draw[mod, rounded corners=\r] (\h*\l+\sp*\ll,0) -- (\h*\l+\sp*\ll,2.3*\l-\sp*\ll) -- (1.5*\h*\l,2.3*\l-\sp*\ll);
            \draw[mod, rounded corners=\r] (\sp*\ll,0) -- (\sp*\ll,\l-\sp*\ll) -- (\h*\l-\sp*\ll,\l-\sp*\ll) -- (\h*\l-\sp*\ll,0);
            \draw[mod, rounded corners=\r] (.5*\h*\l,\l+\sp*\ll) -- (\h*\l-\sp*\ll,\l+\sp*\ll) -- (\h*\l-\sp*\ll,2.3*\l-\sp*\ll) -- (\sp*\ll,2.3*\l-\sp*\ll) -- (\sp*\ll,\l+\sp*\ll) --cycle;
            \draw[speObj, ->-=.55] (1.5*\h*\l,2.3*\l) -- (-.5*\h*\l,2.3*\l);
            \node[mor, opacity=.6, rectangle, inner sep=7.4pt, rounded corners=1pt] at (0,2.3*\l) {};
            \node[mor, opacity=.6, rectangle, inner sep=7.4pt, rounded corners=1pt] at (\h*\l,2.3*\l) {};
            \node[mor, opacity=.6, regular polygon, regular polygon sides=3, inner sep=3.9pt, rounded corners=1pt, rotate=30] at (\h*\l-.04,\l) {};
            \node at (\h*\l,\l) {${\sss #7}$};
            \node[mor, opacity=.6, regular polygon, regular polygon sides=3, inner sep=3.9pt, rounded corners=1pt, rotate=-30] at (.04,\l) {};
            \node at (0,\l) {${\sss #6}$};
        }{}
        \ifthenelse{#8=2}{
            \draw[obj,->-=0.75] (0,\l) --node[pos=0.4,fill=white,rectangle,inner sep=1pt] {${\sss #5}$} (\h*\l,\l);
            \draw[obj,->-=0.68] (0,0) --node[pos=0.28,fill=white,rectangle,inner sep=1pt] {${\sss #3}$} (0,\l);
            \draw[obj,->-=0.89] (0,\l) --node[pos=0.49,fill=white,rectangle,inner sep=1pt] {${\sss #1}$} (0,2*\l);
            \draw[obj,->-=0.68] (\h*\l,0) --node[pos=0.28,fill=white,rectangle,inner sep=1pt] {${\sss #4}$} (\h*\l,\l);
            \draw[obj,->-=0.89] (\h*\l,\l) --node[pos=0.49,fill=white,rectangle,inner sep=1pt] {${\sss #2}$} (\h*\l,2*\l);
            \draw[obj,->-=.68] (\h*\l,-.3*\l-\l) --node[pos=0.28,fill=white,rectangle,inner sep=1pt] {${\sss #1}$}  (\h*\l,-.3*\l);
            \draw[obj,->-=.68] (0,-.3*\l-\l) --node[pos=0.28,fill=white,rectangle,inner sep=1pt] {${\sss #1}$}  (0,-.3*\l);
            \node[mor, rectangle, inner sep=7.4pt, rounded corners=1pt] at (0,-.3*\l) {};
            \node[mor, rectangle, inner sep=7.4pt, rounded corners=1pt] at (\h*\l,-.3*\l) {};
            \node[mor, regular polygon, regular polygon sides=3, inner sep=3.9pt, rounded corners=1pt, rotate=-30] at (.04,\l) {};
            \node[mor, regular polygon, regular polygon sides=3, inner sep=3.9pt, rounded corners=1pt, rotate=30] at (\h*\l-.04,\l) {};
            \draw[speObj, ->-=.55] (1.5*\h*\l,-.3*\l) -- (-.5*\h*\l,-.3*\l);
            \draw[mod, rounded corners=\r] (-\sp*\ll,-.3*\l-\l) -- (-\sp*\ll, -.3*\l-\sp*\ll) -- (-.5*\h*\l,-.3*\l-\sp*\ll);
            \draw[mod, rounded corners=\r] (\h*\l+\sp*\ll,-.3*\l-\l) -- (\h*\l+\sp*\ll, -.3*\l-\sp*\ll) -- (1.5*\h*\l,-.3*\l-\sp*\ll);
            \draw[mod, rounded corners=\r] (\sp*\ll,-.3*\l-\l) -- (\sp*\ll, -.3*\l-\sp*\ll) -- (\h*\l-\sp*\ll,-.3*\l-\sp*\ll) -- (\h*\l-\sp*\ll,-.3*\l-\l);
            \node[mor, opacity=.6, rectangle, inner sep=7.4pt, rounded corners=1pt] at (0,-.3*\l) {};
            \node[mor, opacity=.6, rectangle, inner sep=7.4pt, rounded corners=1pt] at (\h*\l,-.3*\l) {};
            \node at (0,\l) {${\sss #6}$};
            \node at (\h*\l,\l) {${\sss #7}$};
        }{}
    \end{tikzpicture}
}

\newcommand{\entanglementSpectrum}[1]{
    \begin{tikzpicture}
        \clip (-0.95,-7.4) rectangle (7.55,2.1);
        \ifthenelse{#1=1}{
        \def\plotName{\textbf{$\boldsymbol{\mathbb A_4}$ SPT phase}}
        \def\phase{A4_SPT}
        \def\xlimit{44}
        \pgfplotsset{param/.style={xtick={6250,12500,18750,25000,31250},xticklabels={100\,KB,200\,KB,300\,KB,400\,KB,500\,KB}}}
        \newcommand\modelList{
          1/Original/LimeGreen/x,
          2/$\Rep(\mathbb Z_3)$/BurntOrange/x,
          3/$\Rep(\mathbb D_2)$/BlueViolet/x,
          4/$\Rep^\psi(\mathbb D_2)$/Thistle/x,
          7/$\Rep(\mathbb Z_2)$/Goldenrod/x,
          5/$\Rep(\mathbb A_4)$/CornflowerBlue/x,
          6/$\Rep^\psi(\mathbb A_4)$/BrickRed/x}
        }{}
        \ifthenelse{#1=2}{
        \def\plotName{\textbf{$\mathbf{\mathbb A_4}$ symmetric phase}}
        \def\phase{A4}
        \def\xlimit{22}
        \pgfplotsset{param/.style={xtick={6250,12500,18750,25000,31250},xticklabels={100\,KB,200\,KB,300\,KB,400\,KB,500\,KB}}}
        \newcommand\modelList{
          1/Original/LimeGreen/x,
          2/$\Rep(\mathbb Z_3)$/BurntOrange/x,
          3/$\Rep(\mathbb D_2)$/BlueViolet/x,
          4/$\Rep^\psi(\mathbb D_2)$/LimeGreen/x,
          7/$\Rep(\mathbb Z_2)$/Goldenrod/x,
          5/$\Rep(\mathbb A_4)$/CornflowerBlue/x,
          6/$\Rep^\psi(\mathbb A_4)$/BurntOrange/x}
        }{}
        \ifthenelse{#1=3}{
        \def\plotName{\textbf{$\mathbf{\mathbb D_2}$ symmetric phase}}
        \def\phase{D2}
        \def\xlimit{24}
        \pgfplotsset{param/.style={xtick={1250,2500,3750,5000},xticklabels={20\,KB,40\,KB,60\,KB,80\,KB}}}
        \newcommand\modelList{
          1/Original/LimeGreen/x,
          2/$\Rep(\mathbb Z_3)$/BurntOrange/x,
          3/$\Rep(\mathbb D_2)$/BlueViolet/x,
          4/$\Rep^\psi(\mathbb D_2)$/LimeGreen/x,
          7/$\Rep(\mathbb Z_2)$/Goldenrod/x,
          5/$\Rep(\mathbb A_4)$/CornflowerBlue/x,
          6/$\Rep^\psi(\mathbb A_4)$/BurntOrange/x}
        }{}
        \pgfplotsset{plotStyle/.style = {scatter,ymin=-3,ymax=0,xmin=0,xmax=\xlimit,ytick = {-1,-2,-3},xtick = \empty,ticklabel style={font=\scriptsize},major tick length = 2pt,width = 3.7cm,height = 1.6cm,title style = {at={(0.98,0.85)},anchor=north east,font=\footnotesize}},scale only axis,ylabel style = {font = \footnotesize}}
        \pgfplotsset{every axis plot/.append style={only marks,scatter/classes={1={mark=*,NavyBlue},2={mark=*,BurntOrange},3={mark=*,BrickRed},4={mark=*,LimeGreen}},scatter src=explicit symbolic,mark size = 0.7pt}}
        \begin{groupplot}[group style = {group size = {2 by 4}, horizontal sep = 0.15cm, vertical sep = 0.15cm}]
            \nextgroupplot[plotStyle,xticklabel={\empty},title = Original,ylabel = $\log(\lambda)$,ylabel shift = -0.1cm]
            \addplot table [x index = 0, y index = 1, meta index = 2] {data/\phase_1.dat};
            \nextgroupplot[plotStyle,xticklabel={\empty},yticklabel={\empty},title = $\Rep(\mathbb Z_3)$]
            \addplot table [x index = 0, y index = 1, meta index = 2] {data/\phase_2.dat};
            \nextgroupplot[plotStyle,xticklabel={\empty},title = $\Rep(\mathbb D_2)$,ylabel = $\log(\lambda)$,ylabel shift = -0.1cm]
            \addplot table [x index = 0, y index = 1, meta index = 2] {data/\phase_3.dat};
            \nextgroupplot[plotStyle,xticklabel={\empty},yticklabel={\empty},title = $\Rep^\psi(\mathbb D_2)$]
            \addplot table [x index = 0, y index = 1, meta index = 2] {data/\phase_4.dat};
            \nextgroupplot[plotStyle,xticklabel={\empty},title = $\Rep(\mathbb A_4)$,ylabel = $\log(\lambda)$,ylabel shift = -0.1cm]
            \addplot table [x index = 0, y index = 1, meta index = 2] {data/\phase_5.dat};
            \nextgroupplot[plotStyle,xticklabel={\empty},yticklabel={\empty},title = $\Rep^\psi(\mathbb A_4)$]
            \addplot table [x index = 0, y index = 1, meta index = 2] {data/\phase_6.dat};
            \nextgroupplot[plotStyle,xticklabel={\empty},title = {$\Rep(\mathbb A_4)$, rescaled},ylabel = $\log(\lambda)$,ylabel shift = -0.1cm]
            \addplot table [x index = 0, y index = 1, meta index = 2] {data/\phase_8.dat};
            \nextgroupplot[plotStyle,xticklabel={\empty},yticklabel={\empty},title = $\Rep(\mathbb Z_2)$]
            \addplot table [x index = 0, y index = 1, meta index = 2] {data/\phase_7.dat};
        \end{groupplot}
        \node at ($(group c1r1.north)!0.5!(group c2r1.north)+(0,0.3cm)$) {\plotName};
        \begin{axis}[name=param,anchor=north west,at=(group c1r4.south west),yshift = -0.15cm,y dir = reverse, legend columns = 4, ytick={4,5,6},yticklabels={$-4$,$-5$,$-6$},ymin=3,ymax=6,height = 1.6cm,width = 7.55cm,xmin=0,param,scaled ticks=false,ticklabel style={font=\scriptsize},legend style={font=\footnotesize,draw=none,fill=none,/tikz/every even column/.append style={column sep=0.1cm},/tikz/every odd column/.append style={column sep=0.02cm},at={(1,1)},anchor=north east},legend image post style={only marks},major tick length = 2pt,ylabel = $\log(\lambda_{\text{min}})$,ylabel shift = -0.1cm,ylabel style = {font = \footnotesize}]
        \foreach \i/\model/\lineColor/\markType in \modelList{
            \ifthenelse{\i=7}{
            \expandafter\addlegendentry\expandafter{}
            \expandafter\addlegendimage\expandafter{empty legend}
            }{}
            \edef\temp{\noexpand%
            \addplot+[color=\lineColor,mark=\markType,solid,mark size = 1.5pt,sharp plot] table [y index = 0, x index = \i]{data/param_\phase.dat};
            }\temp
            \expandafter\addlegendentry\expandafter{\model}
            }
        \end{axis}
    \end{tikzpicture}
}

\newcommand{\triangleDiagram}[1]{
    \begin{tikzpicture}[scale=0.6,baseline={([yshift=-.5ex]current bounding box.center)}]
    \def\l{7.5};
    \def\ll{4};
    \def\lll{4.6};
    \ifthenelse{#1=1}{
        \node (a) at (0,0) {Symmetry};
        \node (b) at (\l,0) {Bond algebra};
        \node (c) at ({\l*cos(60)},{-.8*\l*sin(60)}) {Quasiparticles};
        \draw[<->] (a) --node[pos=.5, fill=white, rectangle, inner sep=2pt, text=BlueViolet] {${\sss\text{Kinematics}}$} (b);
        \draw[<->] (b) --node[pos=.5, fill=white, rectangle, inner sep=2pt, text=BlueViolet] {${\sss\text{Entanglement}}$} (c);
        \draw[<->] (a) --node[pos=.5, fill=white, rectangle, inner sep=2pt, text=BlueViolet] {${\sss\text{Gapped phase}}$} (c);
    }{}
    \ifthenelse{#1=2}{
        \node (a) at (0,0) {$\mc C$};
        \node (b) at (\ll,0) {$\mc C^*_\mc R$};
        \node (c) at ({\ll*cos(60)},{-.9*\ll*sin(60)}) {$\mc C^*_\mc P$};
        \draw[<->] (a) --node[pos=.5, fill=white, rectangle, inner sep=2pt, text=BlueViolet] {${\sss \mc R}$} (b);
        \draw[<->] (b) --node[pos=.5, fill=white, rectangle, inner sep=2pt, text=BlueViolet] {${\sss \mc Q}$} (c);
        \draw[<->] (a) --node[pos=.5, fill=white, rectangle, inner sep=2pt, text=BlueViolet] {${\sss \mc P}$} (c);
    }{}
    \ifthenelse{#1=3}{
        \node (a) at (0,0) {$\mc C^*_\mc P$};
        \node (b) at (\ll,0) {$\mc C^*_\mc R$};
        \node (c) at ({\ll*cos(60)},{-.9*\ll*sin(60)}) {$\mc C^*_\mc P$};
        \draw[<->] (a) --node[pos=.5, fill=white, rectangle, inner sep=2pt, text=BlueViolet] {${\sss \mc Q}$} (b);
        \draw[<->] (b) --node[pos=.5, fill=white, rectangle, inner sep=2pt, text=BlueViolet] {${\sss \mc Q}$} (c);
        \draw[<->] (a) --node[pos=.5, fill=white, rectangle, inner sep=2pt, text=BlueViolet] {${\sss \mc C^*_\mc P}$} (c);
    }{}
    \ifthenelse{#1=4}{
        \node (a) at (0,0) {$\Vect_{\mathbb A_4}$};
        \node (b) at (\lll,0) {$\Rep(\mathbb A_4)$};
        \node (c) at ({\lll*cos(60)},{-.9*\lll*sin(60)}) {$({\Vect_{\mathbb A_4}})^*_{\mc M(K,\phi)}$};
        \draw[<->] (a) --node[pos=.5, fill=white, rectangle, inner sep=2pt, text=BlueViolet] {${\sss \Vect}$} (b);
        \draw[<->] (b) --node[pos=.5, fill=white, rectangle, inner sep=2pt, text=BlueViolet] {${\sss \Rep^\phi(K)}$} (c);
        \draw[<->] (a) --node[pos=.5, fill=white, rectangle, inner sep=2pt, text=BlueViolet] {${\sss \mc M(K,\phi)}$} (c);
    }{}
    \end{tikzpicture}
}

\begin{document}

\title{Entanglement and the density matrix renormalisation group\\ in the generalised Landau paradigm}

\author{\textsc{Laurens Lootens}}
\email{ll708@cam.ac.uk}
\affiliation{Department of Applied Mathematics and Theoretical Physics, University of Cambridge,\\ Wilberforce Road, Cambridge, CB3 0WA, United Kingdom}
\affiliation{Department of Physics and Astronomy, Ghent University, Krijgslaan 281, 9000 Gent, Belgium}
\author{\textsc{Clement Delcamp}}
\affiliation{Institut des Hautes \'Etudes Scientifiques, Bures-sur-Yvette, France}
\author{\textsc{Frank Verstraete}}
\affiliation{Department of Applied Mathematics and Theoretical Physics, University of Cambridge,\\ Wilberforce Road, Cambridge, CB3 0WA, United Kingdom}
\affiliation{Department of Physics and Astronomy, Ghent University, Krijgslaan 281, 9000 Gent, Belgium}

\begin{abstract}
\noindent
The fields of entanglement theory and tensor networks have recently emerged as central tools for characterising quantum phases of matter. In this article, we determine the entanglement structure of ground states of gapped symmetric quantum lattice models, and use this to obtain the most efficient tensor network representation of those ground states. We do this by showing that degeneracies in the entanglement spectrum arise through a duality transformation of the original model to the unique dual model where the entire dual (generalised) symmetry is spontaneously broken and subsequently no degeneracies are present. Physically, this duality transformation amounts to a (twisted) gauging of the unbroken symmetry in the original ground state. This result has strong implications for the complexity of simulating many-body systems using variational tensor network methods. For every phase in the phase diagram, the dual representation of the ground state that completely breaks the symmetry minimises both the entanglement entropy and the required number of variational parameters. We demonstrate the applicability of this idea by developing a generalised density matrix renormalisation group algorithm that works on (dual) constrained Hilbert spaces, and quantify the computational gains obtained over traditional tensor network methods in a perturbed Heisenberg model. Our work testifies to the usefulness of generalised non-invertible symmetries and their formal category theoretic description for the practical simulation of strongly correlated systems.
\end{abstract}

\maketitle

\noindent
The concept of symmetry breaking forms one of the cornerstones of many-body physics. This was already recognised by Pierre Curie in 1894  who wrote that it is \emph{dissymmetry} that creates all interesting phenomena \cite{curie1894symetrie,castellani2016curie}. Landau formalised this idea, and suggested that different gapped symmetry breaking phases may be characterised by local order parameters that transform non-trivially under the symmetry \cite{landau1936theory,landau1936theory}. In 1971, Kadanoff and Ceva turned the tables and demonstrated that non-local order parameters can be used to characterise gapped symmetric phases \cite{kadanoff1971determination}. In the case of the Ising model, their non-local order parameter is obtained by applying the celebrated Kramers--Wannier duality transformation to the local one \cite{kramers1941statistics}. Such non-local order parameters play a prominent role in the characterization of topological phases of matter \cite{denNijs:1989ntw,PhysRevB.45.304,PhysRevB.85.075125,chen2011classification}: these phases cannot be distinguished from trivial phases by local order parameters, and therefore challenge the standard Landau paradigm.

The modern approach to characterising phases of matter uses the language of entanglement and quantum circuits. Two gapped Hamiltonians with a given symmetry are in the same phase if and only if there exists a symmetry-preserving sub-linear depth quantum circuit that transforms their ground states into each other \cite{bravyi2006lieb,hastings2010localityquantumsystems}. In one dimension, it has been proven that there is an area law for the entanglement entropy for ground states of local gapped Hamiltonians \cite{hastings2007area}. Such states can be efficiently represented in terms of matrix product states (MPS) \cite{Verstraete:2006mdr}, whose (topological) phase can be characterised directly by the transformation properties of the entanglement degrees of freedom under the symmetry \cite{chen2011classification,schuch2011classifying}. Importantly, the MPS description of a ground state can efficiently be obtained using the density matrix renormalisation group (DMRG). These tensor network based algorithms effectively break down the exponential complexity wall for finding ground states of interacting one-dimensional quantum lattice models \cite{PhysRevLett.69.2863,schollwock2011density,cirac2021matrix}. The less entanglement in the ground state, the better these algorithms perform.

While traditional symmetry operators act in an on-site manner, tensor networks enable the definition of more general symmetries on the lattice that encode a correlated action on neighbouring sites. Mathematically, such generalised symmetries \cite{PhysRevLett.98.160409,Bultinck:2015bot,Aasen_2016,Buican:2017rxc,Chang:2018iay,Thorngren:2019iar,Aasen:2020jwb,PhysRevResearch.2.043086,Freed:2022qnc,Schafer-Nameki:2023jdn,Shao:2023gho} are described by a \emph{fusion category} \cite{etingof2016tensor}, and they can be represented explicitly as matrix product operators (MPO) \cite{Sahinoglu:2014upb,Vanhove,10.21468/SciPostPhys.10.3.053,Bridgeman:2022gdx,PRXQuantum.4.020357}. The phases of such systems are classified by a choice of \emph{module category} compatible with those symmetries \cite{Kitaev:2011dxc,Thorngren:2019iar,Komargodski:2020mxz,Thorngren:2021yso,Inamura:2021szw,GarreRubio2023classifyingphases}, and the \emph{generalised Landau paradigm} \cite{PhysRevResearch.2.043086,Gaiotto:2020iye,Apruzzi:2021nmk,Bhardwaj:2023fca,Huang:2023pyk} entails that the inclusion of these generalised symmetries yields a complete classification of gapped phases. Crucially, module categories also classify the different ways a given (generalised) symmetry can be gauged \cite{Frohlich:2006ch,Bhardwaj:2017xup,10.21468/SciPostPhys.8.1.015}. The explicit operators representing those operations on the lattice are again of the MPO form \cite{PRXQuantum.4.020357}, and they map local order parameters to non-local ones, generalising the aforementioned Kramers--Wannier duality. By allowing them to act on the boundary conditions, one can show that these can be lifted to unitary operations and hence preserve the full spectrum of the Hamiltonian \cite{PRXQuantum.5.010338,PhysRevLett.134.130403}.

In this paper, we connect the above concepts of symmetry breaking, duality, and entanglement, and come to the following insight: for every Hamiltonian, one can determine an optimal dual Hamiltonian whose ground states minimise the amount of entanglement. More precisely, consider a Hamiltonian with a (generalised) symmetry and a ground state in a certain phase. By a (twisted) gauging of the remaining symmetries, the corresponding dual Hamiltonian will spontaneously break its dual symmetry completely. Doing so eliminates the degeneracy imposed by the symmetry on the entanglement spectrum, minimising the entanglement compared to all other dual ground states. Additionally, a completely broken symmetry means all variational parameters in the ground state are independent, leading to a large reduction in computational complexity. The ground states in the original theory can be obtained by multiplying the optimal dual ground states with duality operators in the form of an MPO \cite{PRXQuantum.4.020357,PRXQuantum.5.010338}, which reintroduces the multiplicities in the entanglement spectra and enlarges the bond dimension. In the case of twisted gauging, this procedure maps symmetry protected topological phases to trivial symmetry broken ones and vice-versa. 

The only price we pay by considering a dual model is that it may not be defined on a tensor product Hilbert space due to kinematical constraints introduced by the gauging procedure. We overcome this by developing a variant of the DMRG algorithm that directly incorporates these constraints, and demonstrate that all the building blocks for state-of-the-art implementations of tensor networks algorithms are still in place. When applying our method to an ordinary symmetric phase, the variational parameters are effectively the same as in symmetric tensor network methods \cite{mcculloch2002non,singh2010tensor,weichselbaum2012non}. However, by defining Hamiltonians directly in the dual space \cite{PRXQuantum.4.020357,PRXQuantum.5.010338}, our method yields a much simpler way of optimising over those parameters, and furthermore it allows to extend these methods to work in any phase.

\bigskip  
\noindent\textbf{\large Illustrating example} \\
\noindent\textbf{Hamiltonian and gapped phases} \\
\noindent Before introducing the general framework, we demonstrate our formalism with a Heisenberg-like model representing various gapped phases. In spite of this rather simple model hosting a conventional on-site symmetry, the optimal dual models will exhibit generalised symmetries, demonstrating their importance in computational methods. Although we illustrate our approach for a two-site nearest neighbour Hamiltonian, our formalism readily generalises to longer-range Hamiltonians, as well as transfer matrices of classical statistical mechanical models.

Consider an open quantum spin chain of length $L$, and assign to every site $\msf i$ spin-$1$ degrees of freedom with spin operators $(S_\msf i^x,S_\msf i^y, S_\msf i^z)$. The dynamics is governed by the following family of Hamiltonians \cite{PhysRevB.94.045136}:
\begin{equation}
    \label{eq:HamA4}
    \mathbb H = 
    \sum_{\msf i=1}^{L-1} \big(\mathbb h_{\msf i,0} + J_1 \mathbb h_{\msf i,1} + J_2 \mathbb h_{\msf i,2} \big) \, ,
\end{equation}
with coupling constants $J_1,J_2$ and local operators
\begin{align*}
    \mathbb h_{\msf i,0} &:= 
    S_\msf i^x S_{\msf i+1}^x
    + S_\msf i^y S_{\msf i+1}^y
    + S_\msf i^z S_{\msf i+1}^z \, ,
    \\ 
    \mathbb h_{\msf i,1} &:= 
    (S^x_\msf i S^x_{\msf i+1})^2 + (S^y_\msf i S^y_{\msf i+1})^2 + (S^z_\msf i S^z_{\msf i+1})^2 \, , 
    \\
    \mathbb h_{\msf i,2} &:= \{S^x_\msf i,S^y_\msf i\} S^z_{\msf i+1} + \{S^z_\msf i,S^x_\msf i\} S^y_{\msf i+1} + \{S^y_\msf i,S^z_\msf i\} S^x_{\msf i+1} \, .
\end{align*}
While the term $\mathbb h_{\msf i,0}$ defines the spin-1 \emph{antiferromagnetic Heisenberg} model, which is ${\rm SO}(3)$ symmetric, the terms $\mathbb h_{\msf i,1}$ and $\mathbb h_{\msf i,2}$ are perturbations breaking the symmetry down to the finite subgroup $\mathbb A_4 \subset {\rm SO}(3)$ of orientation-preserving symmetries of the tetrahedron.\footnote{Specifically, $\mathbb A_4$ is isomorphic to the semidirect product $\mathbb Z_3 \ltimes \mathbb D_2$, where the cyclic group $\mathbb Z_3$ is generated by any cyclic permutation of $(S^x,S^y,S^z)$, whereas the dihedral group $\mathbb D_2 \cong \mathbb Z_2 \times \mathbb Z_2$ of order four is generated by $e^{i \pi S^x}$ and $e^{i \pi S^z}$.} The spin-1 degrees of freedom transform as the three-dimensional representation of $\mathbb A_4$, which we denote by $\ub 3$.

In the presence of a symmetry $\mathbb A_4$, we distinguish \emph{seven} possible \emph{gapped} phases. Each gapped phase is labelled by a subgroup $H \subseteq \mathbb A_4$ together with a class $[\psi]$ in the second cohomology group of $H$, which classifies its projective representations.
Physically, while $H$ characterises the symmetry preserved within the ground state subspace, a non-trivial $[\psi]$ signals the presence of edge modes that transform projectively under the action of the subsymmetry $H$ \cite{chen2011classification}. Up to isomorphisms, $\mathbb A_4$ counts five subgroups, namely $\mathbb A_4$, $\mathbb D_2$, $\mathbb Z_3$, $\mathbb Z_2$ as well as the trivial one. Out of these subgroups, only $\mathbb A_4$ and $\mathbb D_2$ have non-trivial second cohomology groups, which are both isomorphic to $\mathbb Z_2$, hence seven gapped phases.

\medskip
\noindent\textbf{Dual ground states as matrix product states} \\
\noindent Following ref.~\cite{PRXQuantum.4.020357,PRXQuantum.5.010338}, Hamiltonians dual to eq.~\eqref{eq:HamA4} can be obtained by acting with a duality operator $\mathbb D$ on the local $\mathbb A_4$ symmetric terms of the Hamiltonian as
\begin{equation}
    \mathbb h^{\text{dual}}_{\msf i,n} \circ \mathbb D  =   \mathbb D \circ \mathbb h_{\msf i,n} \, , \q  n \in \{0,1,2\} \, .
\end{equation}  
These operators preserve the algebra generated by the local symmetric operators and thus they preserve 
the spectrum of the Hamiltonian up to degeneracies \cite{PRXQuantum.5.010338,PhysRevLett.134.130403}. In particular, ground state(s) of the original Hamiltonian can be obtained from  ground state(s) of any dual model as $|\psi_\text{g.s.}\rangle = \mathbb D^\dagger |\psi_\text{g.s.}^\text{dual}\rangle$. Note that these dualities typically relate models whose ground state subspaces have different dimensions. By carefully considering the action of the duality on boundary conditions---and, if necessary, introducing ancillary degrees of freedom---these duality transformations can be lifted to unitary matrices \cite{PRXQuantum.4.020357,PhysRevLett.134.130403}. 

The list of possible duality operators $\mathbb D$ at our disposal, and thus the list of possible dual Hamiltonians $\mathbb H^\text{dual}$, is completely determined by the symmetry of the original model. For the symmetry group $\mathbb A_4$, the various dual models are labelled by a choice of subgroup $H \subseteq \mathbb A_4$ together with a class $[\psi]$ in the second cohomology group of $H$, matching the classification of possible gapped phases with a total of seven distinct dual Hamiltonians. The duality associated with the pair $(H,[\psi])$ amounts to the $\psi$-twisted gauging of the subgroup $H$ \cite{Bhardwaj:2017xup,10.21468/SciPostPhys.8.1.015}. This gauging procedure requires the introduction of gauge degrees of freedom, which are labelled by irreducible projective representations of $H$ with respect to $\psi$, on the links between neighbouring sites. These projective representations and their intertwiners are organised into an algebraic structure denoted by $\Rep^\psi(H)$. Moreover, the gauge degrees of freedom are required to satisfy local Gauss constraints, so that dual models typically act on a Hilbert space that is not a tensor product space.

The ground states for these dual models can be parametrised as MPSs, which are states of the form
\begin{equation*}
    |\psi_\text{g.s.}^\text{dual}\rangle = \sum_{i_1,\ldots,i_L} A^{i_1}_{1} A^{i_2}_{2} \cdots A^{i_L}_{L} |i_1 i_2 \ldots i_L\rangle \, .
\end{equation*}
with each $A^i_{\msf i}$ a $\chi_{\msf i-1} \times \chi_{\msf i}$ matrix. The various $\chi_{\msf i}$ are referred to as bond dimensions and are such that $\chi_{0} = \chi_{L} = 1$. Since these ground states satisfy an area law for the entanglement entropy, the bond dimensions $\chi_{\msf i}$ that characterise the amount of entanglement between neighbouring sites do not scale with the system size $L$. These MPS live in a Hilbert space where the physical $\Rep^\psi(H)$-labelled gauge degrees of freedom on the links are duplicated to the two corresponding vertices, so that the Gauss constraints can be imposed at the vertices. This implies that every virtual bond index of the MPS carries an additional label of the corresponding gauge degree of freedom, and0 allows us to label the entanglement spectrum with labels in $\Rep^\psi(H)$. To compute these MPS ground states, we use a generalised DMRG algorithm that manifestly preserves the constraints in the Hilbert space (see Methods). Once we obtain the ground state of a dual model, it can be transformed to a ground state of the original model using the corresponding duality operator, which can itself be parametrised as an MPO of the form
\begin{equation*}
    \mathbb D = \sum_{\substack{i_1,\ldots,i_L \\ j_1,\ldots,j_L}} D^{i_1 j_1} D^{i_2 j_2} \cdots D^{i_L j_L} |i_1 i_2 \ldots i_L\rangle \langle j_1 j_2 \ldots j_L| \, ,
\end{equation*}
with each $D^{ij}$ a $\chi \times \chi$ matrix. The uncontracted matrix indices on site $1$ and $L$ correspond to the different boundary conditions for the duality MPO, which are the additional degrees of freedom required to make these duality operators unitary. Since this MPO itself has a non-trivial bond dimension $\chi$, its action onto an MPS yields generically yields an MPS of larger bond dimension with more entanglement.

\medskip
\noindent\textbf{Numerical results}\\
\noindent
We consider points $(J_1,J_2)=\{(1,1),(-2,-5),(-5,1)\}$ in the phase diagram of the model \eqref{eq:HamA4} that represent the $\mathbb A_4$ SPT phase, the $\mathbb A_4$ symmetric phase and the $\mathbb D_2$ symmetric phase, respectively. For each point, we simulate all seven dual Hamiltonians labelled by $\Rep^\psi(H)$, which also includes the original one. For the three phases, we plot the \emph{entanglement spectra} of the dual ground states coloured by the different objects labelling the gauge degrees of freedom, as well as the memory requirements to reach a specific minimal Schmidt value $\lambda_\text{min}$ serving as a proxy error measure. Our findings are displayed on fig.~\ref{fig:A4_SPT}, \ref{fig:A4_SYM} and \ref{fig:D2_SYM} and analysed below.

\begin{figure}[t]
    \entanglementSpectrum{1}
    \caption{Entanglement spectra of the dual models in the middle of the ground state on 60 sites in the $\mathbb A_4$ SPT phase of the initial model ($J_1 = 1$, $J_2 = 1$). The colour of a Schmidt value indicates the object that labels the corresponding gauge degree of freedom. Bottom: The memory required to store a ground state MPS tensor in the bulk at double precision for a given truncation error $\lambda_{\text{min}}$. The ground state of the $\Rep^\psi(\mathbb A_4)$ model minimizes the entanglement and the number of variational parameters for a fixed truncation error.}
    \label{fig:A4_SPT}
\end{figure}

\medskip \noindent
$\bul$ \emph{$\mathbb A_4$ SPT, fig.}~\ref{fig:A4_SPT}: In this phase, entanglement degrees of freedom of the unique ground state transform as \emph{projective} representations of $\mathbb A_4$. The three irreducible projective representations of $\mathbb A_4$ are two-dimensional, which explains the two-fold degeneracy for every Schmidt value. Comparing the entanglement spectra of the various dual models, we observe that the entanglement is minimised in the $\Rep^\psi(\mathbb D_2)$ and $\Rep^\psi(\mathbb A_4)$ models. Additionally, the ground state of the $\Rep^\psi(\mathbb A_4)$ model requires the least amount of memory, making this model the most efficient one to simulate. Crucially, this dual model possesses a non-invertible $\Rep(\mathbb A_4)$ symmetry, whereby symmetry operators are labelled by representations of $\mathbb A_4$, which is completely broken in the ground state subspace. An important subtlety with a non-invertible symmetry breaking phase is that the different ground states, being related by non-trivial MPOs, can have different entanglement properties, making them distinguishable. In this particular example, the ground state labelled by the 3d irreducible representation requires more entanglement and therefore more variational parameters than the other ground states. In practice, it is possible to avoid this ground state by biasing the initial MPS towards the other less entangled ground states.

All the other dual models admit ground states that preserve some symmetry,  resulting in these cases in degeneracy in the entanglement spectrum as well excessive memory requirements. This is ostensible in the $\Rep(\mathbb D_2)$ model, where every Schmidt value has a four-fold degeneracy due to its ground state preserving a dual $\mathbb A_4$ symmetry, the $\mathbb D_2$ subgroup of which permuting the gauge degrees of freedom that label the Schmidt values, which happen to be labelled by irreducible representations of $\mathbb D_2$. A more subtle case is the $\Rep(\mathbb A_4)$ model, which has a non-invertible $\Rep(\mathbb A_4)$ symmetry that is only partially broken in the ground state subspace. Although the remaining symmetry constrains the entanglement spectrum, the degeneracy is hidden as the ratios between consecutive Schmidt values labelled by the 3d and a 1d irreducible representation are fixed to $\sqrt 3$. We make this manifest by rescaling the spectrum appropriately.

\begin{figure}[t]
    \entanglementSpectrum{2}
    \caption{Top: Entanglement spectra of the dual models in the middle of the ground state on 60 sites in the $\mathbb A_4$ symmetric phase of the original model ($J_1 = -2$, $J_2 = -5$). The colour of a Schmidt value indicates the object that labels the corresponding gauge degree of freedom. Bottom: The memory required to store a ground state MPS tensor in the bulk at double precision for a given truncation error $\lambda_{\text{min}}$. The ground state of the $\Rep(\mathbb A_4)$ model minimises the entanglement and number of variational parameters for a fixed truncation error.}
    \label{fig:A4_SYM}
\end{figure} 

\begin{figure}[t]
    \entanglementSpectrum{3}
    \caption{Top: Entanglement spectra of the dual models in the middle of the ground state on 60 sites in the $\mathbb D_2$ symmetric phase of the original model ($J_1 = -5$, $J_2 = 1$). The colour of a Schmidt value indicates the object that labels the corresponding gauge degree of freedom. Bottom: The memory required to store a ground state MPS tensor in the bulk at double precision for a given truncation error $\lambda_{\text{min}}$. The ground state of the $\Rep(\mathbb D_2)$ model minimizes the entanglement and the number of variational parameters for a fixed truncation error.}
    \label{fig:D2_SYM}
\end{figure}

\medskip \noindent
$\bul$ \emph{$\mathbb A_4$ symmetric, fig.}~\ref{fig:A4_SYM}: In this phase, entanglement degrees of freedom of the unique ground state transform as \emph{linear} representations of $\mathbb A_4$, the three-dimensional irreducible representation $\ub 3$ explaining occurrences of three-fold degeneracy in the entanglement spectrum. Comparing the entanglement spectra of the various dual models, we observe that the entanglement is minimised in the $\Rep(\mathbb A_4)$ model. Additionally, it is the model whose ground state requires the least amount of memory, making this model the most efficient one to simulate. As for the previous phase, this optimal dual model has a non-invertible symmetry $\Rep(\mathbb A_4)$, which also happens to be completely broken in the ground state subspace. Finding the ground state of this optimal dual model is effectively what existing symmetric tensor network methods achieve (see Methods).

All the other dual models admit ground states sharing the same entanglement spectrum as the initial model, differing only in the labelling of the Schmidt values and the improvement in the memory requirements. These dual models possess either invertible or non-invertible symmetries, and the ground states break various amounts thereof, but never is the whole symmetry broken. To verify the absence of hidden degeneracies, we have rescaled the entanglement spectrum of the $\Rep(\mathbb A_4)$ model.

\medskip \noindent
$\bul$ \emph{$\mathbb D_2$ symmetric, fig.}~\ref{fig:D2_SYM}: In this phase, entanglement degrees of freedom of the ground states transform as linear representations of $\mathbb D_2$. Since irreducible representations of $\mathbb D_2$ are all one-dimensional, no additional degeneracy in the entanglement spectrum is enforced, the visible two-fold degeneracies must originate from a hidden symmetry that might involve time reversal combined with a physical on-site action. Comparing the entanglement spectra of the various dual models, we observe that the entanglement is minimised in the initial model, as well as the $\Rep^\psi(\mathbb D_2)$,  $\Rep(\mathbb D_2)$ and $\Rep(\mathbb Z_2)$ models. However, the $\Rep(\mathbb D_2)$ model stands out as requiring the least amount of memory, making this model the most efficient one to simulate. This optimal dual model has an $\mathbb A_4$ symmetry, which happens to be completely broken in the ground state subspace.

The remaining dual models admit ground states showing more entanglement, as a consequence of Schmidt values in the initial ground state becoming degenerate. This is most easily understood in the $\Rep(\mathbb Z_3)$ model that possesses a non-invertible $\Rep(\mathbb A_4)$, which is fully preserved by the unique ground state. The one-dimensional irreducible representations act by permuting gauge degrees of freedom labelled by irreducible representations of $\mathbb Z_3$, enforcing an additional three-fold degeneracy. A similar explanation holds for the degenerate Schmidt values in the $\Rep(\mathbb A_4)$ and $\Rep^\psi(\mathbb A_4)$ models. To verify the absence of hidden degeneracies, we have rescaled the entanglement spectrum of the $\Rep(\mathbb A_4)$ model.

\medskip \noindent
In all the above examples, the dual model that is the most computationally efficient to simulate in terms of memory requirements is always the unique one whose dual symmetry is completely broken in the ground state subspace. This is the model obtained by performing the (possibly) twisted gauging of the symmetry that is preserved within the ground state subspace of the initial model. Performing the ground state computation in the specific dual model obtained by gauging the full symmetry of the Hamiltonian is effectively equivalent (see Methods) to the approach taken by symmetric tensor network methods \cite{mcculloch2002non,singh2010tensor,weichselbaum2012non}. As we have seen however, this is only optimal in the symmetric phase, and imposing too much symmetry requires additional long-range entanglement which adversely affects the performance of the algorithms. By going to the optimal dual model where all symmetry is broken, our methods do not suffer from this and outperform the current state of the art algorithms, both in memory and computational time requirements. Additionally, they are much simpler to implement and uncover the mathematical structure underlying the quasiparticle excitations, as we argue in the next section.

\bigskip
\noindent\textbf{\large General framework}\\
\noindent
The results presented above hold much more broadly. Consider a \emph{generalised} symmetry in a one-dimensional quantum lattice model, i.e., a symmetry whose operators are not necessarily invertible \cite{Aasen_2016,Bhardwaj:2017xup,Buican:2017rxc,Chang:2018iay,Thorngren:2019iar,Aasen:2020jwb,PhysRevResearch.2.043086,Schafer-Nameki:2023jdn,Shao:2023gho}. The symmetry operators take the form of (typically non-local) MPOs \cite{
Vanhove,10.21468/SciPostPhys.10.3.053,Bridgeman:2022gdx,PRXQuantum.4.020357,PRXQuantum.5.010338}. Mathematically, a \emph{finite} generalised symmetry can be modelled by a so-called \emph{fusion category} \cite{Etingof:2002vpd}, extending the group theoretical formalism of ordinary symmetries. Similar to ordinary symmetries, generalised symmetries can be spontaneously broken---as we already witnessed in our examples---and may be gauged provided that there is no 't Hooft anomaly. These possible gaugings are classified by a choice of \emph{module category} over the symmetry fusion category, which is the same classification as the possible gapped phases with respect to such a symmetry \cite{Chang:2018iay,Thorngren:2019iar,Komargodski:2020mxz,Huang:2021zvu,Inamura:2021szw,Bhardwaj:2023idu}. 

Consider a one-dimensional quantum lattice model with a generalised symmetry. Suppose the symmetry is completely broken in the ground state subspace. By gauging the symmetry, which amounts to identifying the corresponding symmetry operators as well as the corresponding symmetry broken states, we obtain a dual model with a dual symmetry that is fully preserved by the unique ground state. Crucially, gauging this dual symmetry recovers the initial model. More generally, there is always a way to gauge the (sub)symmetry that is preserved in the ground state subspace of a model so as to yield a dual model whose dual generalised symmetry is completely broken (see Methods). Practically, this dual model is obtained by extending the approach followed in our series of examples. First, we write the Hamiltonian in terms of tensors that make the generalised symmetry of the model manifest. These tensors satisfy equations generalising 
eq.~\eqref{eq:Fmove}. Then, there is a different set of solutions to these equations, which correspond to the relevant gauging of the preserved subsymmetry, yielding the dual model (see Methods). 

We claim that the optimal way of simulating the phase of a given model amounts to simulating the dual phase of this dual model where the dual symmetry is completely broken, after which we recover the original ground state by acting with the MPO that transmutes the corresponding Hamiltonians into each other. Broadly speaking, the reasoning is that any symmetry translates into constraints amongst the variational parameters so they are not all independent. The associated redundancy unequivocally translates into a suboptimal use of computational resources, as we observed in the examples above.

In the optimal dual phase, at least one of the ground states has the property that the action of any dual symmetry operator on it yields an orthogonal ground state. In this phase, the ground states are in one-to-one correspondence with symmetry operators, and the state having this property is that corresponding to the identity operator. For this maximal symmetry breaking state, all order parameters are strictly local. This follows from the fact that the action of the MPOs representing the dual symmetries map such a ground state into a different one, and hence the expectation value of any non-local string order operator vanishes exponentially in the number of sites on which it acts \cite{haegeman2015shadows}. Conversely, all quasiparticle excitations on top of this maximal symmetry breaking ground state correspond to domain wall excitations. In the case of an infinite (1+1)d lattice model, these excitations are created by the action of the symmetry MPOs on one half of the chain \cite{PhysRevB.85.100408}. As in the usual ansatz for topological excitations in MPS, additional variational degrees of freedom characterising the precise nature of the excitations emerge at the endpoint of the MPO \cite{Zauner_exc,Marien_Fib}. The domain wall excitations of the dual symmetry breaking model are mapped to the quasiparticle excitations of the original model, which can be of a very different nature (spinon, holon, etc.). However, the equivalence between dual theories implies that the fusion category describing the quasiparticle excitations remains the same providing a characterisation of these excitations in any possible gapped phase. Our work reveals the underlying relations between the category theoretic structures describing the properties of a gapped ground state of a (generalised) symmetric Hamiltonian in (1+1)d. Graphically, these relations are summarised in fig.~\ref{fig:comm_diagram_words}.

\begin{figure}
    \centering
    \triangleDiagram{1}
    \caption{The arrows denote relations between the fusion categories that organise the symmetry, the bond algebra generated by the Hamiltonian terms and the quasiparticle excitations. Given an abstract bond algebra of symmetric operators, there are different choices for the kinematical degrees of freedom on which it can be represented. A particular choice then determines the explicit Hamiltonian, and subsequently, its symmetries. Similarly, given a symmetry, there are different gapped phases that a system with such a symmetry can exhibit. A particular choice of phase then determines what the possible quasiparticles are, e.g. domain wall or charge excitations. This diagram means that the composition of these relations is consistent in the sense that the kinematical degrees of freedom together with the phase of the model uniquely specify the structure of the edge modes, or equivalently the entanglement degrees of freedom. We explain this diagram in more detail in fig.~\ref{fig:comm_diagram}.}
    \label{fig:comm_diagram_words}
\end{figure}

\bigskip
\noindent\textbf{\large Discussion}\\
\noindent 
Although fusion categories deal with a finite number of objects, our formalism is more general and also works for continuous symmetries described by Lie groups. In particular, when dealing with a model with an on-site $\text{SU}(2)$ symmetry in the symmetric phase, the required duality MPO boils down to the IRF-vertex transformation \cite{1988CMaPh.118..355P,sierra1997density}, which is a special case of the more general Schur-Weyl duality for $\text{SU}(N)$. Currently, our algorithm only takes advantage of \emph{internal} symmetries, since it exploits an approach to dualities that has been tailored to this type of symmetry. It will be very interesting to generalise this approach to different types of symmetry, such as spatial symmetries and time reversal, which lead to further refinements of phase diagrams.

While we have restricted to gapped phases, we expect that our approach will also be useful to study phase transitions between gapped phases. At these critical points, the ground state no longer admits a description in terms of an MPS with finite bond dimension, and working with such an MPS induces a relevant perturbation to the critical Hamiltonian. By increasing the bond dimension, the strength of this relevant perturbation decreases, and one use this to extrapolate and obtain accurate predictions for critical exponents; this is known as \emph{entanglement scaling} \cite{NISHINO199669,PhysRevB.78.024410}. The nature of the relevant perturbation depends strongly on the symmetries present in the MPS, and by using different dual models we are effectively approaching the critical point from different directions. By combining the entanglement scaling results from these different dual models, we expect an improvement in the accuracy of the predicted critical exponents.

One of the main merits of our method is that it is systematically extensible to higher dimensional models, and to tensor network network algorithms in terms of projected entangled pair states \cite{verstraete2008matrix}. In fact, taking advantage of symmetries in higher dimensions is expected to be even more beneficial than in (1+1)d. Although formally more challenging, many aspects of dualities in two-dimensional quantum lattice models have been worked out \cite{10.21468/SciPostPhys.16.6.143}. Specifically, the relevant tensor network operators are already known for a large class of generalised symmetries \cite{PhysRevX.5.011024,Delcamp:2021szr,Delcamp:2023kew}. In particular, models in (2+1)d with 1-form symmetries admit tensor network representations where these symmetries manifest themselves on the entanglement degrees of freedom, and are therefore robust under perturbations such that they also capture models where the physical higher form symmetry is emergent \cite{haegeman2015shadows}. These virtual symmetries become physical in the double layer transfer matrix, which can then be exploited in the computation of the (1+1)d environment using the methods in this work. We expect this type of dimensional reduction to hold in higher dimensions as well, where the computational gains are even more significant.

\bigskip\bigskip\noindent
\textbf{\large Acknowledgments}\\
\noindent
We would like to thank Jacob Bridgeman, Lukas Devos, Jos\'e Garre-Rubio, Jutho Haegeman, Sukhwinder Singh and Maarten Van Damme for interesting discussions and useful comments. This work has received funding from EOS (grant No. 40007526), IBOF (grant No. IBOF23/064), BOF-GOA (grant No. BOF23/GOA/021). LL is supported by an FWO postdoctoral fellowship (grant No. 12AUN24N) and an EPSRC postdoctoral fellowship (grant No. EP/Y020456/1).

\newpage
\bigskip
\noindent\textbf{\large Methods}\\
\noindent\textbf{Dualities}\\
\noindent
We explain here how to obtain the models dual to \eqref{eq:HamA4} considered in the main text. We encourage the reader to consult ref.~\cite{PhysRevLett.134.130403} for additional details. Firstly, it is crucial to rewrite the local operators entering the definition of the Hamiltonian \eqref{eq:HamA4} in such a way that the symmetry $\mathbb A_4$ is manifest. Invoking a finite group version of the \emph{Wigner--Eckart} theorem, we know that any operator transforming trivially under $\mathbb A_4$ must be expressible as a linear combination of \emph{Clebsch--Gordan} coefficients. The group $\mathbb A_4$ possesses three one-dimensional irreducible representations $\{\ub 0,\ub 1, \ub 1^*\}$ and a single three-dimensional one $\ub 3$ satisfying $\ub 3 \otimes \ub 3 \cong \ub 0 \oplus \ub 1 \oplus \ub 1^* \oplus 2 \cdot \ub 3$. Given three irreducible representations $V_1$, $V_2$ and $V_3$ such that $V_3 \subset V_1 \otimes V_2$, we interpret the intertwining map $V_3 \to V_1 \otimes V_2$ as the tensor 
\begin{equation*}
    \PEPST{vac}{}{}{}{i}{}{}{}{V_1}{V_2}{V_3}{1}
    \hspace{-12pt} \equiv \sum_{\substack{v_1,v_2 \\v_3}} \hspace{-8pt}
    \PEPST{vac}{v_1}{\,v_2}{v_3}{i}{}{}{}{V_1}{V_2}{V_3}{1} \hspace{-8pt}
    | V_1,v_1 \ra \, | V_2,v_2 \ra \, \la V_3,v_3| \, ,
\end{equation*}
where the sums are over basis vectors, and $i$ enumerates the different ways $V_1 \otimes V_2$ decomposes into $V_3$. In this equation, the diagram on the r.h.s. depicts the \emph{Clebsch--Gordan} coefficients valued in $\mathbb C$. In this notation, we can show that the local operators $\mathbb h_{\msf i,n}$, $n \in \{0,1,2\}$, are all of the form 
\begin{equation}
    \label{eq:hamA4_CG}
    \mathbb h_{\msf i,n} \equiv \sum_V\sum_{i,j}h_{n}(V,i,j)
    \, \hamOp{\ub 3}{\ub 3}{\ub 3}{\ub 3}{V}{i}{j}{1} \, ,
\end{equation}
where $h_{n}(V,i,j) \in \mathbb C$. Notice that in this formulation, the state space of a given spin-1 degree of freedom is spanned by $|\ub 3,v\ra$, with $v=1,\ldots,3$.

Importantly, we have the following equality of intertwining maps $V_4 \to V_1 \otimes V_2 \otimes V_3$
\begin{equation}
    \label{eq:Fmove}
    \raisebox{2.3em}{\Fmove{1}{V_1}{V_2}{V_3}{V_4}{V_5}{i}{j}{vac}} \hspace{-20pt}
    =
    \sum_{V_6}
    \sum_{k,l}    
    \big( \F{}^{V_1V_2V_3}_{V_4} \big)_{V_5,ij}^{V_6,kl}
    \hspace{-11pt}
    \raisebox{2.3em}{\Fmove{2}{V_3}{V_2}{V_6}{V_4}{V_1}{k}{l}{vac}} \!\!\! ,
\end{equation}
where the `$F$-symbols' $\big( \F{}^{V_1V_2V_3}_{V_4} \big)_{V_5,ij}^{V_6,kl} \in \mathbb C$ are provided by the \emph{Racah W-coefficients} of $\mathbb A_4$. One can use this identity to show that the structure constants of the algebra generated by the local symmetric operators $\{\mathbb h_{\msf i,n}\}_{\msf i,n}$ only depend on the $F$-symbols and coefficients $\{h_{n}\}_{n}$. Interpreting eq.~\eqref{eq:Fmove} as a tensor network equation, one can ask the following question: Is there another set of tensors, generically depicted as
\begin{equation}
    \label{eq:PEPS}
    \PEPST{mod}{}{}{}{i}{}{}{}{V_1}{V_2}{V_3}{1} \hspace{-16pt}
    \equiv
    \sum_{\substack{m_1,m_2 \\ m_3}}
    \hspace{-14pt}
    \PEPST{mod}{m_1\; }{\;\, m_2}{m_3}{i}{}{}{}{V_1}{V_2}{V_3}{1} \hspace{-12pt}
    |V_1,m_1 \ra |V_2,m_2 \ra \la V_3 ,m_3| \, ,
\end{equation}
also indexed by triplets of representations $\{V_1,V_2,V_3 \subset V_1 \otimes V_2\}$, satisfying eq.~\eqref{eq:Fmove}? Keeping $\{h_n\}_n$ the same, replacing the intertwining maps in eq.~\eqref{eq:hamA4_CG} by these new tensors would result in an isomorphic algebra of local operators, and thus a dual model with Hamiltonian \cite{PRXQuantum.4.020357}
\begin{equation}
    \label{eq:dual_HamA4}
    \mathbb h^{\rm dual}_{\msf i,n} = \sum_V\sum_{i,j}h_{n}(V,i,j)
    \, \hamOp{\ub 3}{\ub 3}{\ub 3}{\ub 3}{V}{i}{j}{2} \, ,
\end{equation}
which can be checked to have the same spectrum \cite{PRXQuantum.5.010338}.
When dealing with a symmetry $\mathbb A_4$, one can find collections of tensors of the form \eqref{eq:PEPS} satisfying eq.~\eqref{eq:Fmove} for any $H \subseteq \mathbb A_4$ and $[\psi] \in H^2(H,{\rm U}(1))$ \cite{10.1155/S1073792803205079}. This is because dualities correspond to (twisted) \emph{gauging} maps of the symmetry $\mathbb A_4$, and there are as many ways to gauge a symmetry as there are ways to spontaneously break it. 
Typically, the dual Hamiltonian acts on a distinct microscopic Hilbert space, which is not necessarily a tensor product space as suggested by the graphical notation. This is consistent with the gauging interpretation, leading to theories with gauge degrees of freedom that satisfy local \emph{Gauss constraints}.  
Given a pair $(H,[\psi])$, degrees of freedom of the resulting dual model are associated with irreducible $\psi$-projective representations \cite{PRXQuantum.4.020357}, and as such we label the model by $\Rep^\psi(H)$. Instead of Clebsch--Gordan coefficients, the tensors are found to evaluate to Racah $W$-coefficients involving linear representations of $G$ and $\psi$-projective representations of the subgroup $H$. Crucially, the initial Hamiltonian can be transmuted into any of its duals via an MPO \cite{PRXQuantum.5.010338}:
\begin{equation}
    \inter{\ub 3}{\ub 3}{\ub 3}{\ub 3}{V}{i}{j}{1}
    = 
    \inter{\ub 3}{\ub 3}{\ub 3}{\ub 3}{V}{i}{j}{2} \, ,
\end{equation}
which is true for any $V$, $i$ and $j$.

\medskip
\noindent\textbf{Generalised DMRG}\\
\noindent
In order to compute the ground states of the various models appearing in the main text, we implemented a version of the \emph{two-site} DMRG algorithm for finite chains with open boundary conditions \cite{white1993density}. The DMRG algorithm is a variational algorithm within the subspace of MPSs, which we recall is a class of wavefunctions that implement the area law of gapped phases at the microscopic level, thereby specifically targeting the physical corner of the total Hilbert space. Briefly, the DMRG algorithm proceeds as follows: The wavefunction being a multilinear function of the variables in all local tensors, the global optimisation problem can iteratively be solved using an \emph{alternating least squares} approach \cite{verstraete2004density}. The two-site version proceeds by blocking two sites together, before solving the combined least-squares problem, and finally using a singular value decomposition in order to split the two-site tensor into two one-site tensors. This two-site version is typically preferred as it allows for an easier redistribution of Schmidt coefficients in the different tensor blocks.

Typically, MPSs are taken to span a subspace of a tensor product space. But, an important feature of the models we consider is that they are typically not defined on a tensor product space. Rather, states need to satisfy some local kinematical constraints; these are the local Gauss constraints mentioned above. Thus, we require MPSs that explicitly enforce these kinematical constraints. For our illustrative example, this is accomplished by considering tensors of the form  
\begin{equation}
\begin{split}
    \label{eq:MPSTen}
    \raisebox{10pt}{\MPSTen{0}} 
    \equiv \!
    \sum_{\substack{V_1,V_2 \\ d_1,d_2,i}}
    \!\!\!  &\raisebox{15pt}{\MPSTen{1}} \!\!
    \\[-4pt]
    &\;\; \cdot  |V_1,d_1 \ra \, | V_1 \ub 3 V_2,i \ra \, \la V_2,d_2 | \, ,
\end{split}
\end{equation}
where $V_1,V_2$ are summed over $\psi$-projective irreducible representations of some subgroup $H \subseteq \mathbb A_4$, $i$ over basis vectors in the space of intertwining maps $V_1 \otimes \ub 3 \to V_2$, while $d_1,d_2$ label the remaining variational degrees of freedom. Notice that both entanglement degrees of freedom labelled by $(V,d)$ and physical degrees of freedom labelled by $(V_1 \ub 3 V_2,i)$ carry gauge degrees of freedom represented by blue lines, which are shared by neighbouring physical degrees of freedom, as suggested by our graphical notation. For instance, the local Hilbert space on two sites is spanned by vectors of the form $|V_1 \ub 3 V_2,i\rangle|V_2 \ub 3 V_3,j\rangle$. Typically, the dimension of the space of intertwining maps $V_1 \otimes \ub 3 \to V_2$ depends on $(V_1,V_2)$, which is incompatible with a tensor product Hilbert space.
By construction, the action of the Hamiltonian \eqref{eq:dual_HamA4} leaves the constrained Hilbert space invariant, and thus explicitly preserves the structure of such an MPS. 

Our algorithm proceeds like the standard two-site DMRG, but all the basic operations are tailored to preserve the kinematical constraints, which amounts to maintaining the structure displayed in eq.~\eqref{eq:MPSTen}. 
First of all, by using block-diagonal basis transformations on the entanglement space, any MPS of the form \eqref{eq:MPSTen} can be brought into the \emph{left canonical form} defined by the condition:
\begin{equation}
    \densityMat{3} \; \stackrel{!}{=} \;  \densityMat{4} \; .
\end{equation}
In left canonical form, the Schmidt values $\lambda$ of the reduced density matrix obtained by tracing out all the sites to the right of the site $\msf i$ are then given by the spectrum of $\rho_\msf i$ defined as
\begin{equation}
    \densityMat{1} \; \equiv \densityMat{2} \; .
\end{equation}
These density matrices are block diagonal, with blocks labelled by $\psi$-projective irreducible representations of the subgroup $H$. As reviewed above, a crucial step of the two-site DMRG algorithm amounts to decomposing the two-site MPS tensor that solves the combined least-squares problem into single-site MPS tensors $A$. Specifically, consider a tensor whose entries are of the form
\begin{equation}
    \label{eq:DoubleTen}
    \raisebox{16pt}{\MPSDoubleTen{1}} \! .
\end{equation}
Keeping the gauge degree of freedom $V_2$ fixed, one considers the matrix $B^{V_2}$ with entries 
\begin{equation}
    [B^{V_2}]_{(V_1,d_1i_1)}^{(V_3,d_3i_2)} :=
    [B^{(V_1 \ub 3 V_2,i_1)(V_2 \ub 3 V_3,i_2)}]_{(V_1,d_1)}^{(V_3,d_3)} \, .
\end{equation}
Performing a singular value decomposition yields a factorisation of the form $B^{V_2} = M^{V_2} \Sigma^{V_2} (N^{V_2})^\dagger$, where $M$ and $N$ are unitary matrices, while $\Sigma^{V_2}$ is a diagonal matrix. The entries of $\Sigma$ are all positive and are referred to as the singular values of $B^{V_2}$. Truncating the singular values to the desired precision $\lambda_\text{min}$ yields the low-rank approximation
\begin{equation*}
    [B^{V_2}]_{(V_1,d_1i_1)}^{(V_3,d_3i_2)} \approx \!
    \sum_{k=1}^{\lambda_\text{min}} [M^{V_2}]_{(V_1,d_1i_1)}^k \, [\Sigma^{V_2}]_k^k \, {[N^{V_2}]^k_{(V_3,d_3i_2)}}^{\!\! *} \, .
\end{equation*}
Repeating this operation for all $\psi$-projective representations $V_2$ of $A$, one finally defines MPS tensors $A_\msf i$ and $A_{\msf i+1}$ 
\begin{equation}
\begin{split}
    [A_\msf i^{(V_1 \ub 3 V_2,i_1)}]_{(V_1,d_1)}^{(V_2,k)}
    &:= 
    [M^{V_2}]_{(V_1,d_1i_1)}^k
    \\
    \text{and} \q [A_\msf {i+1}^{(V_2 \ub 3 V_3,i_2)}]_{(V_2,k)}^{(V_3,d_3)}
    &:=  [\Sigma^{V_2}]_k^k \, {[N^{V_2}]_{(V_3,d_3 i_2)}^k}^{\!\! *} ,
\end{split}
\end{equation}
respectively , so that \eqref{eq:DoubleTen} is approximated by
\begin{equation}
    \raisebox{16pt}{\MPSDoubleTen{2}} \! .
\end{equation}
We can then repeat the same steps for the sites $\msf i+1$ and $\msf i+2$, and keep on sweeping from left to right and then from right to left.  

In our simulations, we initialise the bulk of the MPS with random matrices of a given dimension per block $V_2$, while on the boundary we restrict ourselves to a single one-dimensional block. This corresponds to a choice of boundary condition for the MPO transmuting the Hamiltonian of the model we are simulating into the initial one. In the very specific cases where our algorithm boils to the standard symmetry-preserving DMRG, this amounts to the customary fixing of the total charge sector of the state. Finally, note that our approach is not specific to the two-site DMRG. In particular, uniform MPS algorithms for infinite chains (including VUMPS and `pulling-through' algorithms) can also be implemented \cite{PhysRevB.97.045145,haegeman2017diagonalizing}.

\medskip
\noindent\textbf{Symmetries in tensor networks}\\
\noindent
Below, we clarify the specific scenarios in which our algorithm amounts to using symmetry-preserving tensors. Consider a Hamiltonian with an ordinary symmetry encoded into a (finite) group $G$. Suppose the Hamiltonian is in the $G$ symmetric phase. We claim that the optimal way of simulating this phase via the DMRG algorithm is to simulate the dual model obtained by gauging $G$, before acting with the MPO transmuting the Hamiltonian of the dual model into the initial one. Within our framework, gauging $G$ means that we are dealing with MPS tensors of the form \eqref{eq:MPSTen} where the gauge degrees of freedom depicted by the blue lines are also labelled by irreducible representations of $G$.  The building blocks of the MPO for this duality then evaluate to Clebsch--Gordan coefficients. More precisely, in the case of the model \eqref{eq:hamA4_CG}, where the local Hilbert space is spanned by $|\ub 3,v \ra$, $v=1,\ldots,3$, we have the following identification:
\begin{equation}
    \label{eq:MPO_CG}
     \MPO \equiv \raisebox{13pt}{\symTN{1}} \!  ,
\end{equation}
whereby the MPO tensor acts on the space of intertwining maps $V_1 \otimes \ub 3 \to V_2$. Acting with this MPO then yields an MPS of the form 
\begin{equation}
    \label{eq:dualMPS}
    \raisebox{25pt}{\dualMPS} \, .
\end{equation}
We show that, in this very specific case, our procedure amounts to directly simulating the initial phase using symmetry-preserving tensor networks \cite{mcculloch2002non,singh2010tensor,weichselbaum2012non}. Generally, symmetry-preserving tensor network algorithms exploit a specific expression for the tensors that explicitly enforces the symmetry.   Concretely, consider an MPS in the Hilbert space of the model \eqref{eq:dual_HamA4}. Generically, entanglement degrees of freedom of the MPS tensor live in a vector space $U$. Let us suppose that the MPS tensors are invariant under the action of $\mathbb A_4$. As already exploited earlier, the Wigner--Eckart theorem stipulates that the tensors are expressible as linear combination of Clebsch--Gordan coefficients. More concretely, since the vector space $U$ is equipped with an $\mathbb A_4$ action, it can be decomposed into irreducible representations, i.e. $U \cong \bigoplus_{V} \la U,V\ra V$, where the direct sum is over irreducible representations of $\mathbb A_4$ and $\la V,U \ra \in \mathbb Z_{\geq 0}$. It follows that we can decompose $u=1,\ldots,\dim U$ as $(V,v,d) \equiv (V,v) \otimes (V,d)$, where $v=1,\ldots,\dim V$ and $d=1,\ldots,\la U,V \ra$. Using this notation, the MPS tensors decompose as follows:
\begin{equation}
    \label{eq:symTN}
    \raisebox{15pt}{\symTN{0}} 
    \!\! \equiv \sum_i \stackrel{\symTN{1}}{\symTN{2}} \! ,
\end{equation}
revealing in particular the \emph{sparse block structure} of the tensors. As per our graphical calculus, the matrices on the r.h.s. labelled by $i$ are in fact of the form \eqref{eq:MPSTen}. From a symmetric tensor network viewpoint, this decomposition is  used to target a specific charge sector of the Hilbert space, thereby reducing computational costs, while explicitly enforcing the symmetry \cite{mcculloch2002non,singh2010tensor,weichselbaum2012non}.  
Let us now assume that the MPS is the unique ground state of the $\mathbb A_4$ symmetric phase. In this case, eq.~\eqref{eq:symTN} precisely recovers \eqref{eq:dualMPS} under the identification \eqref{eq:MPO_CG} such that the entanglement degrees of freedom of the MPS ground state of the dual model are labelled by pairs $(V,d)$. One can now verify that our algorithm then produces result that agree with symmetry-preserving DMRG. However, when comparing our algorithm to current state-of-the-art implementations, our approach turns out to be practically much simpler since it does not require the conventional implementation of the recoupling theory for symmetric tensors based on fusion trees.

As commented in the main text, this decomposition is tailored to the symmetric phase, for which entanglement degrees of freedom of the unique ground state transform as linear representations of $\mathbb A_4$. In contrast, this is clearly not suited to the $\mathbb A_4$ SPT phase, for which entanglement degrees of freedom transform as projective representations of $\mathbb A_4$. Indeed, it is well known that using standard symmetric tensor networks in an SPT phase is more costly because it forces the edge modes to transform according to a linear representation, which requires additional long-range entanglement \cite{PhysRevB.91.115145}.

In practice, symmetry-preserving tensor networks have found most of their utility in models with continuous Lie group symmetries such as $\text{SU}(2)$. While we restricted ourselves to a finite group, our results readily generalise to these cases, implying for instance that a ground state preserving an $\text{SU}(2)$ symmetry is most efficiently described in terms of the ground state of a dual model that breaks a dual non-invertible $\Rep(\text{SU}(2))$ symmetry. For this case, the duality transformation recovers the celebrated Schur--Weyl duality. The resulting dual models are the so-called \emph{interaction-round-a-face} models studied by Sierra and Nishino in ref.~\cite{sierra1997density}. The different symmetry-broken ground states being related by the action of non-trivial symmetry MPOs, these ground states do not have the same entanglement, showing the importance of properly initialising the DMRG algorithm to favour the ground states with the least amount of entanglement. Note that the fact that $\text{SU}(2)$ admits an infinite number of irreducible representations is practically addressed by assigning a weight zero to higher spin ones, so that we effectively deal with a finite number of blocks only, just as in standard DMRG exploiting symmetry-preserving tensors. 

\medskip
\noindent\textbf{Mathematical formalism}\\
\noindent
We summarise here the mathematical formalism underpinning the results presented in the main text. More precise mathematical definitions can be found for instance in ref.~\cite{etingof2016tensor}. Consider any local one-dimensional quantum lattice model with a generalised symmetry encoded into a fusion category $\mc C$. 
In the presence of such a generalised symmetry, gapped phases are characterised by a choice of (indecomposable) \emph{$\mc C$-module category}, whose objects label the degenerate ground states of the phase \cite{Thorngren:2019iar,Komargodski:2020mxz}. The same module categories also characterise the different ways to gauge (sub)symmetries of the model. 
After performing the gauging operation associated with a $\mc C$-module category $\mc M$, the dual symmetry of the resulting model is encoded into the so-called \emph{Morita dual} $\mc C^*_\mc M$ of $\mc C$ with respect to $\mc M$. The fusion category $\mc C^*_\mc M$ is defined to be the category $\Fun_\mc C(\mc M,\mc M)$ of $\mc C$-module endofunctors of $\mc M$ \cite{MUGER200381,etingof2016tensor}, the fusion structure being provided by the composition of $\mc C$-module functors. Crucially, $\mc M$ is also a $\mc C^*_\mc M$ module category and we have ${(\mc C^*_\mc M)}^*_\mc M \simeq \mc C$. In words, there is always a way to gauge a subsymmetry of $\mc C^*_\mc M$ so as to recover the initial model.

Let us examine this gauging operation in practice. For conciseness, we focus on nearest-neighbour Hamiltonians, but longer range interactions can be accommodated just as easily. As for the example studied in the main text, it is crucial to write the Hamiltonian in such a way that the generalised symmetry $\mc C$ is manifest.
Under some mild mathematical assumptions about the symmetry MPOs \cite{10.21468/SciPostPhys.10.3.053,Molnar:2022nmh}, it follows from a generalised Wigner--Eckart theorem \cite{Bridgeman:2022gdx} that any local symmetric operator is expressible in terms of generalised Clebsch--Gordan coefficients. Specifically, given a $\mc C$-symmetric Hamiltonian of the form $\mathbb H = \sum_{\msf i=1}^{L-1}\sum_n \mathbb h_{\msf i,n}$, the local operators can always be put in the form
\begin{equation}
    \label{eq:Ham_gen}
    \mathbb h_{\msf i,n} \equiv \sum_{\{Y\}}\sum_{i,j} h_n(\{Y\},i,j)
    \, \hamOp{Y_1}{Y_2}{Y_3}{Y_4}{Y_5}{i}{j}{2} \, ,
\end{equation}
in terms of tensors evaluating to these generalised Clebsch--Gordan coefficients. The graphical notation mimics that of eq.~\eqref{eq:dual_HamA4} and encodes in particular the fact that the Hilbert space of a model with a generalised symmetry is generically not a tensor product space. More concretely, there is a (possibly not unique) choice of $\mc C$-module category $\mc R$ such that local operators can be expressed as \eqref{eq:Ham_gen} where $\{Y\}$ label objects in $\mc C^*_\mc R$, and the generalised Clesbch--Gordan coefficients are given by the so-called \emph{module associator} of $\mc R$, as a $\mc C^*_\mc R$-module category (see \cite{10.21468/SciPostPhys.10.3.053,PRXQuantum.4.020357,PRXQuantum.5.010338} for details). 

It follows from the defining axioms of the $\mc C^*_\mc R$-module category $\mc R$ that the tensors in eq.~\eqref{eq:Ham_gen} satisfy an analogue to eq.~\eqref{eq:Fmove}:
\begin{equation}
    \label{eq:FmoveG}
    \raisebox{2.3em}{\Fmove{1}{Y_1}{Y_2}{Y_3}{Y_4}{Y_5}{i}{j}{mod}} \hspace{-20pt}
    =
    \sum_{Y_6}
    \sum_{k,l}    
    \big( \F{}^{Y_1Y_2Y_3}_{Y_4} \big)_{Y_5,ij}^{Y_6,kl}
    \hspace{-11pt}
    \raisebox{2.3em}{\Fmove{2}{Y_3}{Y_2}{Y_6}{Y_4}{Y_1}{k}{l}{mod}} \!\!\! ,
\end{equation}
where the `$F$-symbols' $\big( \F{}^{Y_1Y_2Y_3}_{Y_4} \big)_{Y_5,ij}^{Y_6,kl} \in \mathbb C$ enters the definition of the fusion category $\mc C^*_\mc R$. Together with the complex coefficients $\{h_n\}_n$, these $F$-symbols provide the structure constants of the algebra generated by the local symmetric operators $\{\mathbb h_{\msf i,n}\}_{\msf i ,n}$. Within this framework, performing a gauging operation simply amounts to picking a different $\mc C^*_\mc R$-module category $\mc R'$. This means replacing the tensors in eq.~\eqref{eq:Ham_gen} by a new set of tensors that now evaluate to generalised Clebsch--Gordan coefficients given by the module associator of $\mc R'$. Crucially, this new set of tensors still satisfy eq.~\eqref{eq:FmoveG}, so the algebra of local symmetric operators generated by \eqref{eq:Ham_gen} is isomorphic to the initial one. The dual symmetry of the resulting model is then encoded into the fusion category ${(\mc C^*_\mc R)}^*_{\mc R'}$. Similarly to the examples discussed in the main text, Hamiltonians associated with different choices of $\mc C^*_\mc R$-module categories can be transmuted into each other via an MPO. Mathematically, such an operator is described by an object in the category $\Fun_{\mc C^*_\mc R}(\mc R,\mc R')$ of $\mc C^*_\mc R$-module functors from $\mc R$ to $\mc R'$, in such a way that the building blocks of the MPO are provided by the module structure of such a functor \cite{PRXQuantum.5.010338}. Moreover, this category of module functors has the structure of a module category $\mc M$ over the symmetry category $\mc C \simeq \Fun_{\mc C^*_\mc R}(\mc R,\mc R)$ via composition of module functors. We identify this $\mc C$-module category as that encoding the gauging operation corresponding to changing the $\mc C^*_\mc R$-module category $\mc R$ into $\mc R'$, in such a way that the dual symmetry is encoded into $\mc C^*_\mc M \simeq {(\mc C^*_\mc R)}^*_{\mc R'}$. Note that for many fusion categories of interest, indecomposable module categories have been classified. Importantly, the data of the module functors needed to describe the MPO intertwiners relating the different dual models can be obtained as a representation theoretic problem \cite{10.21468/SciPostPhys.13.2.029,Bridgeman:2022gdx}, which numerically can be reduced to a linear algebra problem. 

Let us now suppose that the initial $\mc C$-symmetric model, which is defined with respect to the $\mc C^*_\mc R$-module category $\mc R$, is in the phase associated with the $\mc C$-module category $\mc P$. Our goal is to find a dual model whose dual symmetry is completely broken in the ground state subspace. To achieve this, we must understand how to relate the phase of the dual model to the phase of the initial model, which is not immediate given that the symmetry structures differ. To this end, we can think of the $\mc C$-module category $\mc P$ as $\Fun_{\mc C_\mc R^*}(\mc R, \mc Q)$ for some $\mc C_\mc R^*$ module category $\mc Q$. Indeed, since $\mc C \simeq \Fun_{\mc C_\mc R^*}(\mc R, \mc R)$, it does define a $\mc C$-module category via composition of module functors. Since any $\mc C$-module category can be constructed in this way \cite{etingof2016tensor}, it follows that $\mc P$ uniquely fixes $\mc Q$.
The action of a duality operator $\Fun_{\mc C_{\mc R}^*}(\mc R,\mc R')$ on the phase $\mc P$ can now be obtained via composition of module functors, such that the new phase is encoded into $\Fun_{\mc C_{\mc R}^*}(\mc R',\mc Q)$, which is indeed a module category over the dual symmetry ${(\mc C^*_\mc R)}^*_{\mc R'} = \Fun_{\mc C_{\mc R}^*}(\mc R',\mc R')$.
The optimal dual model, whose dual symmetry is completely broken, is obtained when the module category that describes the dual phase is given by the dual symmetry itself. From the above, it is clear that this amounts to choosing $\mc R' = \mc Q$ so that $\mc Q$ encodes the degrees of freedom.
Consider a variational ground state MPS in the constrained Hilbert space of the optimal dual model with degrees of freedom in $\mc Q$, which maximally breaks the dual symmetry. The remaining symmetry breaking ground states can be obtained by acting with symmetry MPOs labelled by simple objects in $\mc C_{\mc P}^*$ of the form
\begin{equation}
    \symMPO \, ,
\end{equation}
where the individual tensors are determined by the data of the relevant $\mc C_{\mc R}^*$ module functors \cite{PRXQuantum.5.010338}. 
Finally, these dual ground states can be mapped to ground states of the original model, which is defined with respect to the $\mc C^*_\mc R$-module category $\mc R$, by acting with a duality MPO:
\begin{equation}
    \raisebox{25pt}{\dualMPSG} \, ,
\end{equation}
Conversely, the duality operators to the optimal model are encoded into $\Fun_{\mc C^*_\mc R}(\mc R,\mc Q)$, which is equivalent to the $\mc C$-module category $\mc P$ encoding the phase of the original model. The various fusion categories and the Morita equivalences between them are summarised in fig.~\ref{fig:comm_diagram}.

\begin{figure}[t]
    \triangleDiagram{2} $\longmapsto$ \triangleDiagram{3}
    \caption{The action of the symmetry $\mc C$ depends on a choice of module category $\mc R$, which in turn fixes a fusion category $\mc C^*_\mc R$ that governs the algebra of symmetric Hamiltonians. The phase of this Hamiltonian is given by a module category $\mc P$ over $\mc C$, which determines the fusion category $\mc C^*_\mc P$ describing the quasiparticle excitations. Combining these two module categories we obtain $\mc Q = \Fun_\mc C(\mc R,\mc P)$, which describes the entanglement degrees of freedom of the optimal tensor network description of the ground state. Indeed, by dualising and replacing $\mc R$ by $\mc Q$, we end up in the maximal symmetry breaking phase $\mc C^*_\mc P$ of the dual symmetry $\mc C^*_\mc P$.}
    \label{fig:comm_diagram}
\end{figure}

\begin{figure}[t]
    \triangleDiagram{4} 
    \caption{In our illustrating example, the symmetry is $\mathbb A_4$, represented in an on-site manner on a Hilbert space without gauge degrees of freedom, with the algebra of symmetric operators governed by the fusion category of representations of $\mathbb A_4$. A given phase is encoded into a module category $\mc M(K,\phi)$, such that the different ground states are labelled by cosets in $G/K$ and the remaining symmetry $K$ acts $\phi$-projectively. The excitations can be computed as $({\Vect_{\mathbb A_4}})^*_{\mc M(K,\phi)}$, and by composing the arrows in this diagram, we find that the entanglement degrees of freedom are described by $\phi$-projective representations of $K$.}
    \label{fig:comm_diagram_ex}
\end{figure}

To conclude, we revisit our example in light of this general formalism. When dealing with an ordinary (invertible) symmetry $\mathbb A_4$, the corresponding fusion category $\mc C$ is the category $\Vect_{\mathbb A_4}$ of $\mathbb A_4$-graded vector spaces. The different ways to gauge subsymmetries of $\mathbb A_4$ are provided by $\Vect_{\mathbb A_4}$-module categories, which are known to be classified by pairs $(A,[\psi])$ as defined in the main text \cite{10.1155/S1073792803205079}. In particular, choosing $A = G$ and $\psi =1$ amounts to the (untwisted) gauging of $A$, and the corresponding $\Vect_{\mathbb A_4}$-module category is equivalent to the category $\Vect$ of complex vector spaces. The Morita dual $(\Vect_{\mathbb A_4})^*_{\Vect}$ of $\Vect_{\mathbb A_4}$ with respect to $\Vect$ can be checked to be equivalent to the category $\Rep(\mathbb A_4)$ of representations of $\mathbb A_4$, in agreement with our results. When writing the Hamiltonian as in eq.~\eqref{eq:HamA4}, we are choosing the $\Rep(\mathbb A_4)$-module category $\mc R = \Vect$ such that the module associator boils down to the ordinary Clebsch--Gordan coefficients of $\mathbb A_4$. We obtain the various dual models by choosing different $\Rep(\mathbb A_4)$-module categories. Specifically, the dual model resulting from the $\psi$-twisted gauging of $H$ is obtained by choosing the module category $\mc R' = \Rep^\psi(H)$ of $\psi$-projective representations of $H$, and the dual symmetry is encoded into ${(\Rep(\mathbb A_4))}^*_{\Rep^\psi(H)}$.

We now suppose that the initial model is in the phase associated with the $\Vect_{\mathbb A_4}$-module category $\mc M(K,\phi) := \Fun_{\Rep(\mathbb A_4)}(\Vect,\Rep^\phi(K))$. The $\mathbb A_4$ SPT, $\mathbb A_4$ symmetric, and $\mathbb D_2$ symmetric phases considered in the main text are obtained by choosing $\Rep^\phi(K)$ to be equal to $\Rep^\psi(\mathbb A_4)$, $\Rep(\mathbb A_4)$, and $\Rep(\mathbb D_2)$, respectively. The optimal model is always found to be that given by $\mc R' = \Rep^\phi(K)$, which amounts to performing a $\phi$-twisted gauging of $K$, as predicted. The relevant Morita equivalences are summarised in fig.~\ref{fig:comm_diagram_ex}.

Let us also shed light on some of the suboptimal simulations: For instance, consider the $\mathbb A_4$-symmetric phase and the dual model labelled by $\mc R' = \Rep^\psi(\mathbb D_2)$; the symmetry is found to be ${(\Rep(\mathbb A_4))}^*_{\Rep^\psi(\mathbb D_2)} \simeq \Vect_{\mathbb A_4}$, while the dual phase is that associated with the $\Vect_{\mathbb A_4}$-module category $\Fun_{\Rep(\mathbb A_4)}(\Rep^{\psi}(\mathbb D_2), \Rep(\mathbb A_4)) \simeq \Vect$. Together with the fact that $\Rep^\psi(\mathbb D_2)$ is equivalent to $\Vect$ as category, this explains why the numerical results were the same for this dual model as for the initial one. In a similar vein, in the $\mathbb A_4$ SPT phase, the dual model obtained by choosing $\mc R'=\Rep(\mathbb D_2)$ has a ${(\Rep(\mathbb A_4))}^*_{\Rep(\mathbb D_2)} \simeq \Vect_{\mathbb A_4}$ symmetry; the dual phase is associated with the $\Vect_{\mathbb A_4}$-module category $\Fun_{\Rep(\mathbb A_4)}(\Rep(\mathbb D_2), \Rep^\psi(\mathbb A_4))$, which is equivalent to $\Vect$ as a category, meaning that the whole symmetry is preserved, in agreement with our numerical results. Finally, let us examine the $\mathbb D_2$ symmetric phase and the dual model obtained by choosing the $\Vect_{\mathbb A_4}$-module category $\mc R' = \Rep(\mathbb Z_3)$; the dual symmetry ${(\Rep(\mathbb A_4))}^*_{\Rep(\mathbb Z_3)} \simeq \Rep(\mathbb A_4)$ is also fully preserved since the module category over it is $\Fun_{\Rep(\mathbb A_4)}(\Rep(\mathbb Z_3),\Rep(\mathbb D_2))$, which happens to be equivalent to $\Vect$, in agreement with our numerical results.

\bigskip

\end{document}